\documentclass[onecolumn,prb,amsmath,amssymb,showpacs,superscriptaddress]{revtex4-1}
\usepackage{graphics}
\usepackage{graphicx}
\usepackage{epstopdf}
\usepackage{subfigure}
\usepackage{psfrag}
\usepackage{color}
\usepackage{bm}
\usepackage{natbib}
\usepackage{enumerate}

\def\k{{\bm k}}
\def\x{{\bm x}}

\def\c{{\bm c}}
\def\s{{\bm s}}

\def\v{{\bm v}}

\def\dpart#1#2{{\dfrac{\partial #1}{\partial #2}}}
\def\be{\begin{equation}}
\def\ee{\end{equation}}

\def\sinc{{\mbox{sinc}}}

\begin{document}
 \title{Phonon Boltzmann equation-based discrete unified gas kinetic scheme for multiscale heat transfer}
  \author{Zhaoli Guo}
 \email[Email:]{zlguo@mail.hust.edu.cn}
 \affiliation{State Key Laboratory of Coal Combustion, Huazhong University of Science and Technology, Wuhan 430074, China}
 \author{Kun Xu}
 \email[Email:]{makxu@ust.hk}
 \affiliation{Department of Mathematics, Hong Kong University of Science and Technology, Clear Water Bay, Hong Kong, China}

\begin{abstract}
Numerical prediction of multiscale heat transfer is a challenging problem due to the wide range of time and length scales involved. In this work a discrete unified gas kinetic scheme (DUGKS) is developed for heat transfer in materials with different acoustic thickness based on the phonon Boltzmann equation.
With discrete phonon direction, the Boltzmann equation is
discretized with a second-order finite-volume formulation, in which the time-step is fully
determined by the Courant-Friedrichs-Lewy (CFL) condition. The scheme has the asymptotic preserving (AP) properties for both diffusive and ballistic regimes, and can present accurate solutions in the whole transition regime as well. The DUGKS is a self-adaptive multiscale method for the capturing of local transport process. Numerical tests for both heat transfers with different Knudsen numbers are presented to validate the current method.
\end{abstract}

\maketitle
\section{Introduction}
Many emerging nanostructures involve semiconductors and dielectrics, in which
phonon transport is the main mechanism for heat transfer. Heat transfer process in systems with such nanostructures usually involves multiple temporal and spatial scales, \cite{ref:Review0,ref:reviewMinnich2015,ref:ChenBook} and it is a challenging problem to develop efficient numerical methods that are applicable to different transport regimes. Owing to the breakdown of Fourier law at small time and spatial scales, and the high computational requirement of microscopic molecular dynamics, the phonon Boltzmann transport equation (BTE) is regarded to be able to provide a good base for developing numerical methods for multiscale heat transfer when the wave effect of phonon is negligible. Actually, many numerical schemes have been proposed to solve the BTE in previous studies, \cite{ref:Larsen2010} including the stochastic Monte-Carlo (MC) method \cite{ref:MC-Peter,ref:MC-Maz,ref:MC-Hadj2009,ref:MC-Hadj2011} and the deterministic discrete ordinates method (DOM) coupled with finite-difference, finite-volume, or finite-element discretization of spatial space. \cite{ref:NumerFD,ref:DOM-FV,ref:Ali2014,ref:Ye2015,ref:DOM-FE} The lattice Boltzmann method (LBM), which was originally developed for continuous fluid flows \cite{ref:LBMBook}, was also applied to phonon transport. \cite{ref:LBM1,ref:LBM2,ref:LBM3,ref:LBM4,ref:LBM5}

Generally, the MC method follows a time-splitting algorithm, namely, the dynamics of a simulated particle is decoupled into advection and scattering processes, and thus the time step used is less than the relaxation time, and the grid size is less than the phonon mean-free-path. \cite{ref:MC-Zuck} Consequently, the computational costs of MC method are expensive in the acoustic thick regime, which prohibit its applications for multiscale problems with diffusive region, although it can be quite efficient for ballistic transport. It is also noted that an improved MC method has been developed recently by simulating only the deviation from equilibrium such that the variance can be efficiently reduced in simulating systems with small temperature variations. \cite{ref:MC-Hadj2009} In the DOM method, the transient and advection terms in the BTE are usually discretized with techniques that are adopted in computational fluid dynamics (CFD), such as upwind (step) and central (diamond) finite-difference schemes, or finite-volume schemes with upwind interpolations. These CFD techniques may introduce significant artificial diffusions (low-order schemes) or numerical instability (high-order schemes). \cite{ref:Larsen2010} Regarding the LBM for phonon transport, although it has been applied to some nano and multiscale problems, \cite{ref:LBM1,ref:LBM2,ref:LBM3,ref:LBM4,ref:LBM5} some studies have shown that LBM may yield unphysical predictions. \cite{ref:LBM-DOMa,ref:LBM-DOMb}

Recently, a finite-volume discrete unified gas kinetic scheme (DUGKS) for molecule flows ranging from continuum to rarefied regimes has been developed \cite{ref:DUGKS-2013,ref:DUGKS-2015}, which has high accuracy and outstanding robustness. The nice asymptotic preserving (AP) properties also remove the restriction on the time step by the relaxation time that exists in other kinetic methods with direct discretization of the kinetic equation. Furthermore, the finite-volume formulation enables the DUGKS to handle problems with complex geometries. \cite{ref:DUGKS-Zhu1} Some comparative studies suggest that the DUGKS has better performances over the LBM even for continuum flows. \cite{ref:Wang,ref:DUGKS-Zhu2} In this work, we will extend the DUGKS to phonon transport to construct an efficient method for the whole multiscale heat transport process ranging from diffusive to ballistic regimes.

The remainder of the paper is organized as follows. Section II gives a brief introduction of the phonon BTE, and the DUGKS for the BTE is described in Sec. III. Some numerical simulations are carried out in Sec. IV to test the scheme, and finally a brief summary is given in Sec. V.

\section{Phonon Boltzmann Transport Equation}
In a rigid crystalline solids, the atomic vibrations from equilibrium positions can be quantized as quasi-particles known as phonons, and the system can be considered as a domain filled with a phonon gas. The angular frequency $\omega$ of a phonon is related to the wave number $\bm{k}\in R^3$ through certain dispersion relations $\omega=\omega_{p}(\bm{k})$, with different polarizations or modes of the phonon. The phonon transport can be described by the Boltzmann transport equation in the regime as the wave effect of phonon is negligible \cite{ref:reviewMinnich2015},
\begin{equation}
\label{eq:BTE}
\dpart{f_{p}}{t}+\bm{v}_{p}\cdot\nabla f_{p} = Q_{p},
\end{equation}
where $f_{p}=f_{p}(\x, \k, \bm{s},t)$ (or $=f_{p}(\x, \omega, \bm{s},t)$) is the distribution function dependent on wave number $\bm{k}$ (or frequency $\omega$), polarization $p$, direction $\bm{s}$, and position $\bm{x}$ as well as time $t$;  $\bm{v}_{p}=\partial{\omega}/\partial{\bm{k}}$ is the group velocity with which the phonon of polarization $p$ travels. The term on the right hand side, $Q_{p}$, represents the rate of change of $f_{p}$ due to scattering interactions. Usually the scattering is very complicated and a tractable model widely used is the relaxation time approximation,
\begin{equation}
Q_{p}=-\dfrac{1}{\tau_{p}}\left[f_{p}-f_{p}^{eq}\right],
\end{equation}
where $\tau_{p}$ is the relaxation time, $f_{p}^{eq}$ is the equilibrium distribution of phonons and follows the Bose-Einstein distribution,
\begin{equation}
f_{p}^{eq}=\dfrac{1}{\exp\left(\hbar\omega/k_BT\right)-1},
\end{equation}
with $\hbar$ being the Planck's constant divided by $2\pi$ and $k_B$ the Boltzmann constant, respectively, and $T$ is the temperature defined later. The  effective relaxation time $\tau_{p}$ usually depends on temperature and frequency, and can be estimated using the Matthiessen's rule if the individual scattering processes are independent of each other, \cite{ref:Review0,ref:Review2014}
\begin{equation}
\dfrac{1}{\tau_{p}}=\dfrac{1}{\tau_U} + \dfrac{1}{\tau_N} + \dfrac{1}{\tau_b} + \dfrac{1}{\tau_i} + \cdots,
\end{equation}
where the relaxation times appearing on the right hand side are those due to the umklapp (U) and normal (N) phonon-phonon scatterings, boundary scattering, impurity scattering, etc. With the effective relaxation time, one can define the Knudsen number of the system, $\mbox{Kn}=\lambda_0/l_0$, where $l_0$ is the characterize length of the system, and $\lambda_0=v_0\tau_0$ is the phonon mean free path with $v_0$ being a typical value of the phonon group velocity and $\tau_0$ a typical value of the relaxation time.

The total energy and the heat flux can be defined from the phonon distribution function, \cite{ref:Ziman}
\begin{equation}
E=\sum_{\k,p}{\hbar\omega(\k)f_{p}(\k)}=\sum_{p}{\int_{4\pi}\int{\hbar\omega f_p(\omega) D_p(\omega) d\omega}}d{\Omega},
\end{equation}
\begin{equation}
\bm{q}=\sum_{\k,p}{\hbar\omega(\k)\bm{v}_{p}(\k)f_{p}(\k)}=\sum_{p}{\int_{4\pi}\int{\hbar \omega\bm{v}_p} f_p(\omega)D_p(\omega) d\omega}d{\Omega},
\end{equation}
where $D_p(\omega)$ is the density of state, and ${\Omega}$ is the solid angle. The temperature $T$ of the system can be obtained from $T=E/C$, with $C$ being the volume specific heat capacity.

Even with the relaxation time approximation, the BTE is still very difficult to be solved due to the high dimensionality. A number of tractable models have emerged to reduce the complex, such as gray model, semi-gray model, non-gray model, and phonon radiative transfer model.\cite{ref:Review0,ref:Majumda1993} To illustrate the essence of our numerical method clearly without loss of generality, we will consider the gray model in the present work.
This simplified model assumes phonons of all polarizations and frequencies are same, and the BTE \eqref{eq:BTE} is expressed in terms of the phonon energy density $e''(\x,\s,t)$ \cite{ref:Review0},
\begin{equation}
\label{eq:Boltzmann-BGK}
\dpart{e''}{t}+\v\cdot\nabla e''=Q\equiv -\dfrac{1}{\tau}\left[e''-e^{eq}\right],
\end{equation}
where $\v=v_g\s$ is the group velocity with $v_g$ being the magnitude, $\tau$ is the singlet relaxation time, and $e''$ is the reduced distribution function for energy density,
\begin{equation}
e''(\x,\s,t)=\sum_p\int{\hbar\omega f_p(\omega) D_p(\omega)\, d\omega }.
\end{equation}
Obviously, the total phonon energy $E$ and heat flux $\bm{q}$ can be determined from $e''$,
\begin{equation}
E=\int_{4\pi}{e''(\x,\s,t)\, d\Omega},\quad \bm{q}=\int_{4\pi}{\v e''(\x,\s,t)\, d\Omega}.
\end{equation}
The equilibrium distribution function $e^{eq}$ is just the angular average of the total energy,
\begin{equation}
e^{eq}(\x,t)=\dfrac{1}{4\pi}\int_{4\pi}{e''(\x,\s,t)\, d\Omega}=\dfrac{E}{4\pi}.
\end{equation}

The gray model employs a single phonon group speed $v$ in all directions and a single relaxation time $\tau$ independent of polarization and frequency. Despite the simple formulation, the gray model can provide some insightful predictions on the phonon transport behaviors with acceptable accuracy. \cite{ref:DOM-FE,ref:Ye2015} In the diffusive limit ($\mbox{Kn}\to 0$), it can be shown that the solution of the kinetic equation \eqref{eq:Boltzmann-BGK} is determined by its average $E$ that obeys the diffusion equation (see Appendix A for details),
\begin{equation}
\label{eq:MacroEq}
\dpart{E}{t}=\nabla\cdot\left(\kappa\nabla E\right),
\end{equation}
where the thermal conductivity is given by
\begin{equation}
\label{eq:kappa}
\kappa = \dfrac{1}{3} v_g^2 \tau.
\end{equation}

\section{Numerical scheme}
\subsection{Updating rule in finite-volume formulation}
Now we present the construction of the discrete unified gas kinetic scheme (DUGKS) for phonon transport based on Eq. \eqref{eq:Boltzmann-BGK}. First, the continuous solid angle domain $\Omega$ is discretized into $N$ discrete angles using the discrete-ordinates method (DOM) based on certain spherical quadratures, and correspondingly we obtain $N$ discrete directions $\s_i$. The accuracy of the quadrature employed is required to ensure the exact evaluation of the angular moments of the distribution function up to certain orders, such as
\begin{subequations}
\begin{equation}
\sum_{\alpha=1}^N{w_\alpha e''(\s_\alpha)}=\int_{4\pi}{e''(\s)d\Omega} = E,
\end{equation}
\begin{equation}
\sum_{\alpha=1}^N{w_\alpha e^{eq}(\s_\alpha)}=\int_{4\pi}{e^{eq}(\s)d\Omega} = E,
\end{equation}
\begin{equation}
\sum_{\alpha=1}^N{w_\alpha \v_{\alpha} e''(s_\alpha)}=\int_{4\pi}{\v e''(\s)d\Omega} =\bm{q},
\end{equation}
\begin{equation}
\sum_{\alpha=1}^N{w_\alpha \v_{\alpha} e^{eq}(s_\alpha)}=\int_{4\pi}{\v e^{eq}(\s)d\Omega} =0,
\end{equation}
\begin{equation}
\sum_{\alpha=1}^N{w_\alpha \v_{\alpha} \v_{\alpha} e^{eq}(s_\alpha)}=\int_{4\pi}{\v \v e^{eq}(\s)d\Omega} =\dfrac{v_g^2}{3}E\bm{I},
\end{equation}
\end{subequations}
where $w_\alpha$ and $\s_{\alpha}$ are the weights and the discrete angles of the corresponding spherical quadrature, $\v_\alpha=v_g\s_\alpha$ is the discrete group velocity, and $\bm{I}$ is the second-order unit tensor. These requirements suggest that the weights and the discrete angles satisfies
\begin{equation}
\label{eq:w}
\sum{w_{\alpha}}=4\pi, \quad \sum{w_{\alpha}\v_\alpha}=\bm{0},\quad \sum{w_{\alpha}\v_\alpha\v_\alpha}=\dfrac{4\pi}{3}\bm{I}.
\end{equation}
 With the discrete directions, the BTE \eqref{eq:Boltzmann-BGK} can then be expressed as
\begin{equation}
\label{eq:BGK-phi}
\dpart{\phi_\alpha}{t}+\v_\alpha\cdot\nabla\phi_\alpha=Q_\alpha\equiv-\dfrac{1}{\tau}\left(\phi_\alpha-\phi_\alpha^{eq}\right),
\end{equation}
for $\alpha=1, 2, \cdots, N$, where $\phi_\alpha(\x,t)=e''(\x,s_\alpha,t)$ and $\phi_\alpha^{eq}(\x,t)=e^{eq}(\x,s_\alpha,t)$. The total energy density and heat flux are then evaluated from the discrete distribution function for the phonon energy,
\begin{equation}
E=\sum_{\alpha=1}^N{w_\alpha\phi_\alpha},\quad \bm{q}=\sum_{\alpha=1}^N{w_\alpha \v_\alpha\phi_\alpha},
\end{equation}

\begin{figure}
\includegraphics[width=0.45\textwidth]{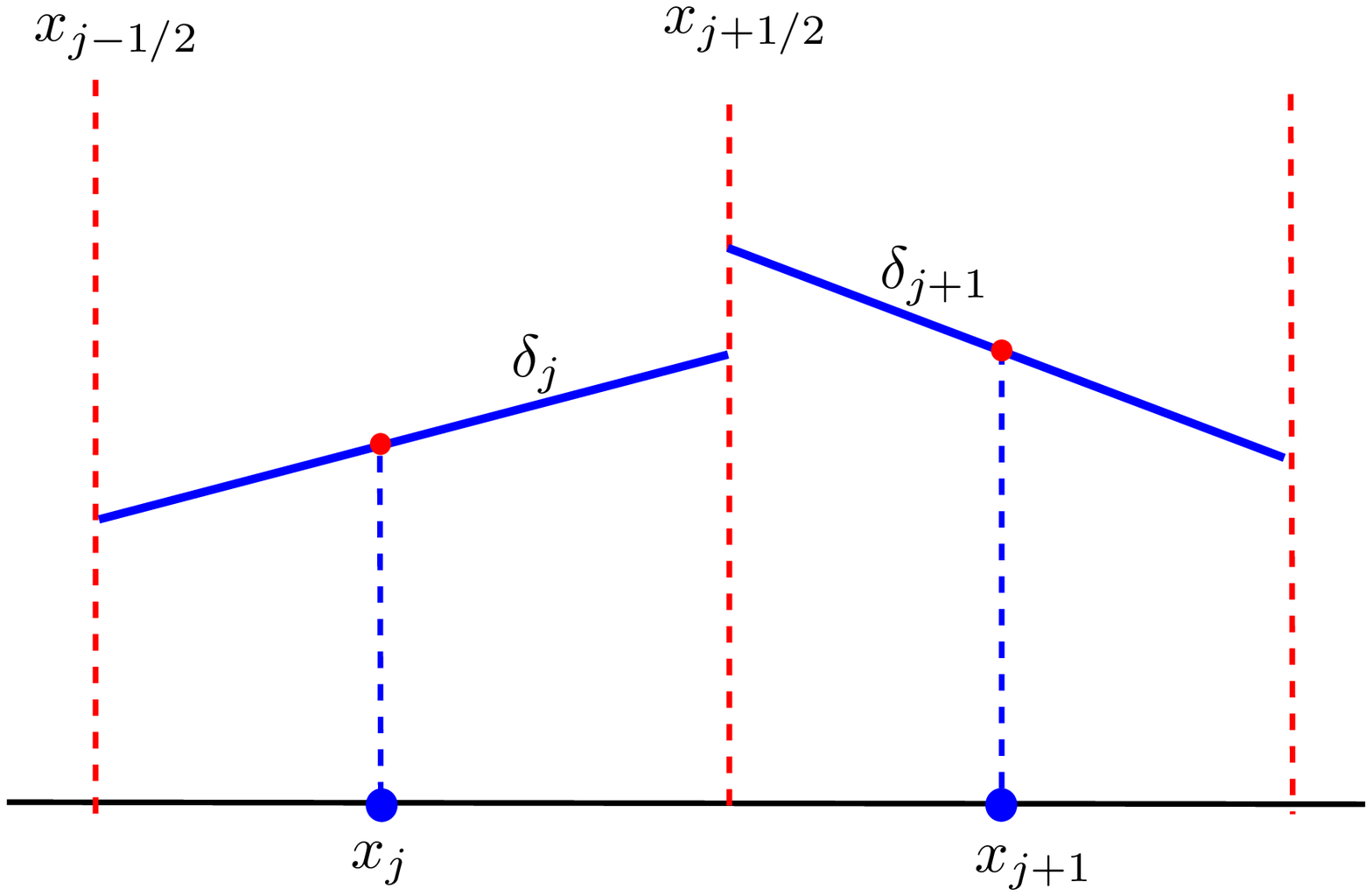}
\includegraphics[width=0.45\textwidth]{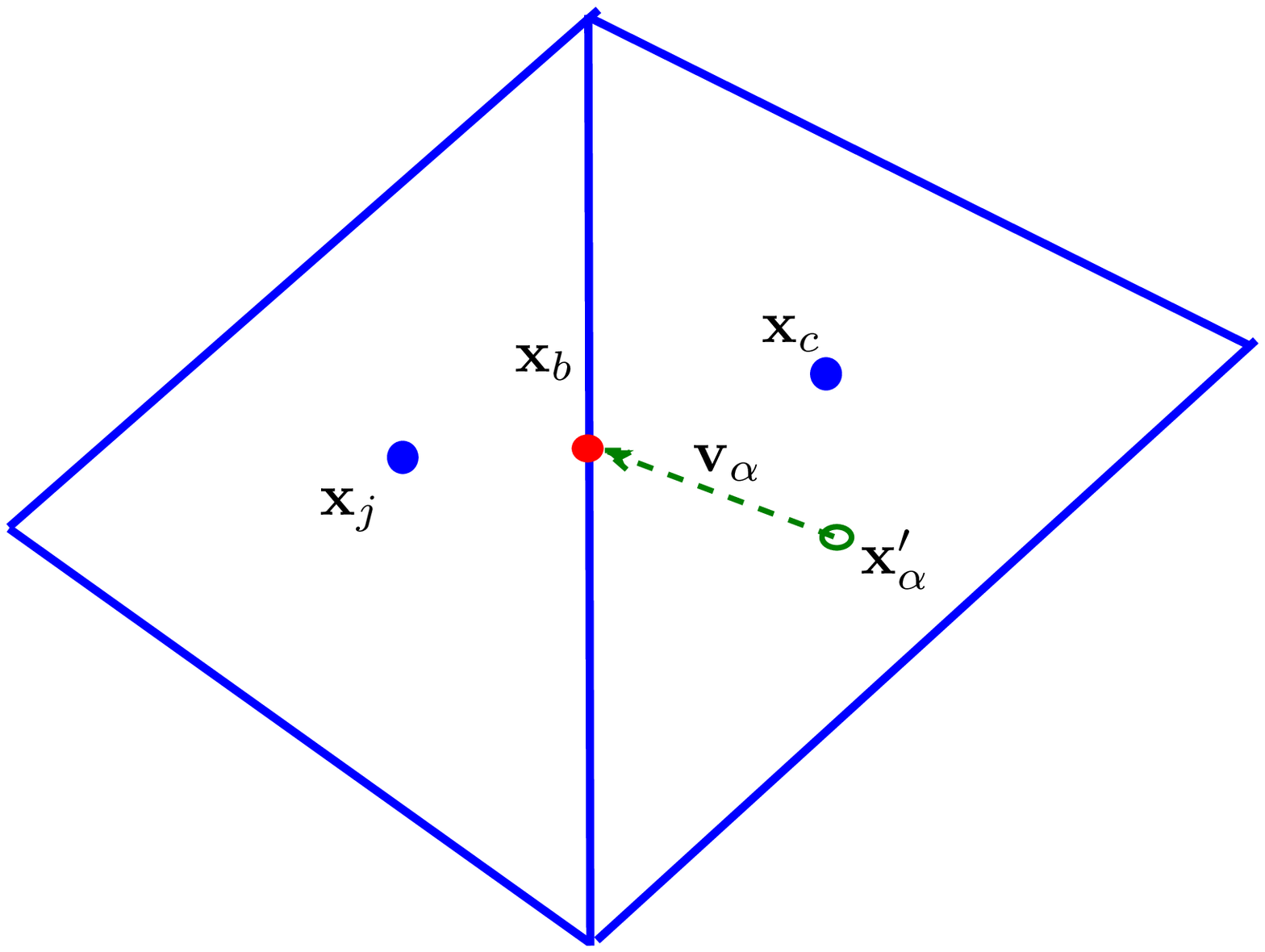}
\caption{Schematic of a 1D (a) and 2D (b) cell geometry.}
\label{fig:Cell}
\end{figure}

The DUGKS method developed here is a finite-volume scheme for solving the BTE \eqref{eq:BGK-phi}, in which the spatial domain is divided into a set of control
volumes. A one-dimensional (1D) and two-dimensional (2D) diagrams are shown in Fig. \ref{fig:Cell} as an example. Integrating Eq. (\ref{eq:BGK-phi}) in the volume $V_j$ centered at
$\x_j$ from time $t_n$ to $t_{n+1}=t_n+\Delta t$ leads to the following balance equation,
\begin{equation}
\label{eq:update_phi}
 \phi_{\alpha j}^{n+1}-\phi_{\alpha j}^n + \dfrac{\Delta t}{|V_j|}{\bm{F}_{\alpha j}^{n+1/2}}=\dfrac{\Delta t
}{2}\left[Q_{\alpha j}^{n+1}+Q_{\alpha j}^n\right],
\end{equation}
where we have used the trapezoidal quadrature for the time integration of the collision term, and the mid-point rule for the flux term. $|V_j|$ is the volume of cell $j$. Here,
\begin{equation}
\label{eq:Flux}
 {\bm{ F}}_{\alpha j}^{n+1/2}=\sum_{\x_b \in{\cal N}_j}{(\v_\alpha\cdot\bm{n}_b) \phi_{\alpha}(\bm{x}_b, t_{n+1/2})S(\x_b)}
\end{equation}
is the flux across the interfaces of cell $j$, where ${\cal N}_j$ is the set consisting of the centers of cell interfaces, $\bm{n}_b$ is the
outward unit normal vector at $\x_b$ of an interface, and $S(\x_b)$ is the corresponding interface area. In Eq. \eqref{eq:update_phi}, $\phi_{\alpha j}$ and
$Q_{\alpha j}$ are the cell-averaged values of the distribution
function and collision term, respectively,
\begin{equation}
\phi_{\alpha j}=\dfrac{1}{|V_j|}\int_{V_j}{\phi_{\alpha}(\x,t)\, d\x}, \quad Q_{\alpha j}=\dfrac{1}{|V_j|}\int_{V_j}{Q_{\alpha}(\x,t)\, d\x}.
\end{equation}

The scheme (\ref{eq:update_phi}) is implicit due to the inclusion of $Q_{\alpha}^{n+1}$ and the flux at half-time step $\bm{F}_{\alpha}^{n+1/2}$.
Like in the original DUGKS for gas flows, \cite{ref:DUGKS-2013,ref:DUGKS-2015} the implicitness of the collision term can be removed by introducing a new
distribution function defined as
\begin{equation}
\label{eq:phi-tilde}
\tilde{\phi}_{\alpha}=\phi_{\alpha}-\dfrac{\Delta t}{2}Q_{\alpha}=\dfrac{2\tau+\Delta t
}{2\tau} \phi_{\alpha} -\dfrac{\Delta t}{2\tau}\phi_{\alpha}^{eq},
\end{equation}
or
\begin{equation}
\phi_{\alpha}=\dfrac{2\tau} {2\tau+\Delta t }\tilde{\phi}_{\alpha}+\dfrac{\Delta
t}{2\tau+\Delta t }\phi_{\alpha}^{eq}.
\end{equation}
Then Eq. (\ref{eq:update_phi}) can be rewritten in terms of $\tilde{\phi}_{\alpha}$ as
\begin{equation}
\label{eq:update_phiA}
 \tilde{\phi}_{\alpha j}^{n+1}=\tilde{\phi}_{\alpha j}^{+,n} - \dfrac{\Delta t}{|V_j|}{\bm{F}}_{\alpha j}^{n+1/2},
\end{equation}
where
\begin{equation}
\label{eq:phi-tilde+}
\tilde{\phi}_{\alpha j}^+=\dfrac{2\tau-\Delta t}{2\tau+\Delta
t}\tilde{\phi}_{\alpha j}+\dfrac{2\Delta t}{2\tau+\Delta t}\phi_{\alpha j}^{eq}.
\end{equation}

Note that the discrete scattering operator conserves energy, i.e, $\sum{w_{\alpha}Q_\alpha}=0$,
Therefore, we can track the evolution of $\tilde{\phi}_{\alpha}$ instead of
$\phi_{\alpha}$, and from the definition of $\tilde{\phi}_{\alpha}$, the energy and heat flux can be computed as
\begin{equation}
E=\sum_{\alpha=1}^N{w_{\alpha}\tilde{\phi}_{\alpha}}, \quad \bm{q}=\dfrac{2\tau}{2\tau+\Delta t}\tilde{\bm{q}}, \quad \mbox{with}\quad \tilde{\bm{q}}=\sum_{\alpha=1}^N{w_{\alpha}\v_{\alpha}\tilde{\phi}_{\alpha}},
\end{equation}
where we have made use the fact that $\sum{w_{\alpha}\v_{\alpha}{\phi}^{eq}_{\alpha}}=\int_{4\pi}{\v\phi^{eq}\,d\Omega}=0$.

\subsection{Flux evaluation based on discrete characteristic solution}
\label{subsec:Flux}
Now we discuss how to evaluate the cell interface flux at the half time-step, $\bm{F}_{\alpha j}^{n+1/2}$. To this end,
we integrate Eq. \eqref{eq:BGK-phi} from $t_n$ to $t_n+h$ (here $h=\Delta t/2$ is the half time step) along the characteristic line with the end point ($\x_b$) locating at the center of the cell interface (see Fig. \ref{fig:Cell}),
\begin{equation}
\label{eq:phi-face}
 \phi_{\alpha}\left(\x_b,t_{n}+h\right)-\phi_{\alpha}\left(\x_b-\v_{\alpha}h,t_{n}\right) =\dfrac{h
}{2}\left[Q_{\alpha}\left(\x_b,t_{n}+h\right)+Q_{\alpha}\left(\x_b-\v_{\alpha}h, t_{n}\right)\right],
\end{equation}
where the trapezoidal rule is again used to evaluate the scattering term. The implicitness in this equation can be removed by introducing another auxiliary distribution function $\bar{\phi}_\alpha$,
\begin{equation}
\label{eq:bar-phi}
\bar{\phi}_{\alpha}=\phi_{\alpha}-\dfrac{h}{2}Q_{\alpha}=\dfrac{2\tau+h
}{2\tau} \phi_{\alpha} -\dfrac{h}{2\tau}\phi_{\alpha}^{eq},
\end{equation}
or
\begin{equation}
\label{eq:fbar2f}
 \phi_{\alpha}=\dfrac{2\tau} {2\tau+h }\bar{\phi}_{\alpha}+\dfrac{h}{2\tau+h
}\phi_{\alpha}^{eq}.
\end{equation}
Then we can obtain from Eq. (\ref{eq:phi-face}) that
\begin{equation}
\label{eq:phi-bar}
 \bar{\phi}_{\alpha}\left(\x_b,t_{n}+h\right)=\bar{\phi}_{\alpha}^+(\x_b-\v_{\alpha}h,t_{n}),
\end{equation}
with
\begin{equation}
\label{eq:phi-bar+}
\bar{\phi}_{\alpha}^+=\dfrac{2\tau-h}{2\tau}\phi_{\alpha}+\dfrac{h}{2\tau}\phi_{\alpha}^{eq}.
\end{equation}
In order to determine $\bar{\phi}_{\alpha}^+(\x_b-\v_{\alpha}h,\s_{\alpha},t_{n})$, we assume that the distribution function is a piecewise linear function in the cell at which $\x_{\alpha}'=\x_b-\v_{\alpha}h$ locates, say the cell centered at $\x_c$ (refer to Fig. \ref{fig:Cell}). Then we can obtain that
\begin{equation}
\label{eq:smooth_phi+}
 \bar{\phi}_{\alpha}^+(\x', t_{n})=
\bar{\phi}_{\alpha}^+(\x_c, t_{n})-(\x_{\alpha}'-\x_c)\cdot\bm{\delta}_c\bar{\phi}_{\alpha}^{+,n},
\end{equation}
where $\bm{\delta}_c\bar{\phi}_{\alpha}^{+}$ is the slope of the distribution function $\bar{\phi}_{\alpha}^{+}$ in the cell centered at $\x_c$, which can be constructed smoothly or using certain numerical limiters. For instance, in 1D case as sketched in Fig. \ref{fig:Cell}, the slope in cell $j$ can be determined by the central-difference,
\begin{equation}
\delta_j\bar{\phi}_{\alpha j}^+=\dfrac{x_{j+1}-x_j}{x_{j+1}-x_{j-1}}s_1 + \dfrac{x_{j}-x_{j-1}}{x_{j+1}-x_{j-1}}s_2,
\end{equation}
for smooth problems, or by the van Leer limiter \cite{ref:vanLeer},
\begin{equation}
\delta_j\bar{\phi}_{\alpha j}^+=\left[\mbox{sgn}(s_1)+\mbox{sgn}(s_2)\right]\dfrac{|s_1||s_2|}{|s_1|+|s_2|},
\end{equation}
for problems with discontinuities, where
\begin{equation}
s_1=\dfrac{\bar{\phi}_{\alpha,j}^+ - \bar{\phi}_{\alpha, j-1}^+}{x_j-x_{j-1}},\quad s_2=\dfrac{\bar{\phi}_{\alpha, j+1}^+ - \bar{\phi}_{\alpha, j}^+}{x_{j+1}-x_{j}}.
\end{equation}
It can be seen that in smooth region where $s_1\approx s_2$, the slopes determined by the two methods are similar. For multi-dimensional case, the slope in each direction can be determined as described above.

Based on Eqs. (\ref{eq:phi-bar}) and (\ref{eq:smooth_phi+}), we can obtain that
\begin{equation}
\label{eq:A_barphi_face}
 \bar{\phi}_{\alpha}(\x_b, t_{n}+h)=\bar{\phi}_{\alpha}^+(\x_c, t_{n})-(\x_{\alpha}'-\c_c)\cdot\bm{\delta}_c\bar{\phi}_{\alpha}^{+,n},
\end{equation}
from which we can obtain the energy at the cell interface,
\begin{equation}
\label{eq:A_Eface}
E(\x_b,t_n+h)=\sum_{\alpha}{\bar{\phi}_{\alpha}(\x_b, t_n+h)},
\end{equation}
where we have again used the energy conservative property of the discrete scattering operator.
Then the equilibrium distribution function $\phi_{\alpha}^{eq}(\x_b,t_n+h)$ can be obtained, and the original distribution function can be
extracted from  $\bar{\phi}_{\alpha}(\x_b,t_{n}+h)$ according to Eq. \eqref{eq:fbar2f},
\begin{equation}
\label{eq:originalFb}
\phi_{\alpha}(\x_b,t_{n}+h)=\dfrac{2\tau} {2\tau+h
}\bar{\phi}_{\alpha}(\x_b,t_{n}+h)+\dfrac{h}{2\tau+h }\phi_{\alpha}^{eq}(\x_b,t_n+h).
\end{equation}
With the known distribution function $\phi_{\alpha}$ at cell interface at the half time step, the flux
$\bm{F}_{\alpha j}^{n+1/2}$ can be evaluated according to Eq. (\ref{eq:Flux}), and the cell-averaged distribution functions $\tilde{\phi}_{\alpha}$ at the new time $t_{n+1}$ can be obtained according to Eq. \eqref{eq:update_phiA}.

\subsection{Algorithm}
Summarizing the updating rule for the cell-averaged distribution functions and the interface flux, the DUGKS consists of the following two equations,
\begin{equation}
\label{eq:center-phi}
 \tilde{\phi}_{\alpha j}^{n+1}=\tilde{\phi}_{\alpha j}^{+,n} - \dfrac{\Delta t}{|V_j|}\sum_{\x_b\in {\cal N}_j}\v_{\alpha}\phi_{\alpha}^{n+1/2}(\x_b),
\end{equation}
\begin{equation}
\label{eq:face-phi}
\phi_{\alpha}^{n+1/2}(\x_b)=\dfrac{2\tau} {2\tau+\Delta t/2
}\left[\bar{\phi}_{\alpha}^{+,n}(\x_c)+(\x'_{\alpha}-\x_c) \cdot\bm{\delta}_c\bar{\phi}_{\alpha}^{+,n}\right]+\dfrac{\Delta t/2}{2\tau+\Delta t/2 }\phi_{\alpha}^{eq,n+1/2}(\x_b),
\end{equation}
where
\begin{equation}
\tilde{\phi}_{\alpha j}^{+,n}=\tilde{\phi}_{\alpha j}^n+\dfrac{2\Delta t}{2\tau+\Delta t}\left[\phi_{\alpha j}^{eq,n}-\tilde{\phi}_{\alpha j}^n\right],
\end{equation}
\begin{equation}
\label{eq:A_barf+}
\bar{\phi}_{\alpha j}^{+,n}=\tilde{\phi}_{\alpha j}^n+\dfrac{3\Delta t/2}{2\tau+\Delta t}\left[\phi_{\alpha j}^{eq,n}-\tilde{\phi}_{\alpha j}^n\right].
\end{equation}
Note that we have made use of Eqs. \eqref{eq:phi-tilde} and \eqref{eq:phi-bar+} in the derivation of Eq. \eqref{eq:A_barf+}.
In practical computations, $\tilde{\phi}_{\alpha}^+$ can be calculated from $\bar{\phi}_{\alpha}^+$,
\begin{equation}
\label{eq:A_tildef+}
 \tilde{\phi}_{\alpha}^+=\dfrac{4}{3}\bar{\phi}_{\alpha}^+
-\dfrac{1}{3}\tilde{\phi}_{\alpha}.
\end{equation}

The calculation procedure of the DUGKS at time step $t_n$ can be listed as follows:
\begin{enumerate}[(i)]
  \item Flux evaluation:
\begin{itemize}
    \item Compute the auxiliary distribution functions
    $\bar{\phi}_{\alpha j}^{+,n}$ according to Eq. \eqref{eq:A_barf+} and its slope in each cell;
    \item Compute the original cell interface distribution function
    $\phi_{\alpha}^{n+1/2}(\x_b)$ according to Eq. \eqref{eq:face-phi}, where $\phi_{\alpha}^{eq,n+1/2}$ is evaluated based on $E^{n+1/2}$ given by Eq. \eqref{eq:A_Eface}.
\end{itemize}
  \item Update of Cell-averaged distribution functions:
\begin{itemize}
    \item Compute the auxiliary distribution functions $\tilde{\phi}_{\alpha j}^{+,n}$
    computed according to Eq. \eqref{eq:A_tildef+};
    \item Update the distribution functions $\tilde{\phi}_{\alpha j}^{n+1}$ in all
    cells via Eq. \eqref{eq:center-phi}.
\end{itemize}

\end{enumerate}

\section{Analysis of the DUGKS}
\subsection{Numerical diffusion}
Artificial diffusion can significantly deteriorate the simulation accuracy of a numerical scheme for the BTE. We now analyze the numerical diffusion of the proposed DUGKS by analyzing the accuracy of the reconstructed cell-interface distribution functions.
First it is noted that the exact solution of the BTE \eqref{eq:BGK-phi} at cell interface center $\x_b$ can be written as
\begin{equation}
\label{eq:ana}
  \phi_{\alpha,e}^{n+1/2}(\bm x_b) = \phi^n_{\alpha}(\x_b - h\v_\alpha) + \int_{0}^{h}Q_{\alpha}\left(\x_b - (h-t')\v_\alpha, t^n+t')\right)\,dt'.
\end{equation}
The first and second terms on the right hand side represents the free transport and scattering processes, respectively.
In the DUGKS, the trapezoidal rule is used to approximate the integral of the scattering term, and the approximation error in this term is $O(h^3)$; For the first term on the right hand side, it is approximated by assuming the distribution function is a linear function in the cell, and the error is of order $O(\Delta x^2)$. Therefore, the overall accuracy of the reconstructed distribution function at a cell interface in DUGKS is $O(\Delta x^2)+O(\Delta t^3)$, and the numerical diffusion will also be of this order since the heat flux is the first-order angular moment of the reconstruction distribution function.

Note that the scattering term in Eq. \eqref{eq:ana} itself is of order $\Delta t$, so if we neglect the scattering term totally (i.e. only the free flight process is considered), like the classical first-order upwind (step) scheme, second-order upwind or central interpolation (diamond) scheme, the overall accuracy of the reconstructed distribution function will be  $O(\Delta x^n)+O(\Delta t)$, where the number $n$ depends on the employed interpolation rule. Therefore, although the use of high-order interpolations can reduce numerical diffusion from spatial discretization, it is no help to reduce numerical diffusion from the scattering integration, which is of order $O(\tau)$. On the other hand, it is known from Eq. \eqref{eq:kappa} that the physical diffusion coefficient is proportional to the relaxation time $\tau$. This suggests that in order to control the numerical diffusion to avoid false diffusion, it is required that $\Delta t \ll \tau$. For problems in near ballistic regime, this is not a problem since $\tau$ is relatively large. However, for diffusive and near diffusive problems, this becomes a rather severe limitation. Therefore, for those explicit BTE solvers that use direct interpolations, the small time step is required not only by the numerical stability condition, but also by the accuracy requirement. This also explains why some implicit BTE solvers could produce large false diffusions even the computation is stable. On the other hand, the numerical diffusion from the discretization of the scattering term in present DUGKS is of order $O(\Delta t^3)$, which can release greatly the restriction on time step by the accuracy requirement.

\subsection{Asymptotic preserving property}
The Asymptotic preserving (AP) property is important for a kinetic scheme in modeling multiscale transport. A kinetic scheme is AP given that \cite{ref:AP-Xu,ref:AP-Mieu} (i) the time step $\Delta t$ is not limited by the relaxation time $\tau$ for any Knudsen number, and (ii) the scheme is consistent with the free transport kinetic equation as $\mbox{Kn}\to \infty$, and consistent with the continuum equation as $\mbox{Kn} \to 0$.

Regarding Point (i), as discussed in the above subsection, the restriction on the time step by accuracy requirement can be much released due to the coupling in the treatment of the scattering and transport processes. Furthermore, the implicitness in the treatment of the collision terms with the trapezoidal rule in both the evolution of the cell-center distribution function [Eq. \eqref{eq:update_phi}] and the reconstruction of the cell-interface distribution  [Eq. \eqref{eq:phi-face}] suggests that the restriction on the time step by numerical stability due to the relaxation time $\tau$ can be removed, too. Therefore, the constraint on the time step of DUGKS due to the free flight process can be ensured by the Courant-Friedrichs-Lewy (CFL) condition,
\begin{equation}
\Delta t =\beta \dfrac{\Delta x_{min}}{v_g},
\end{equation}
where $\Delta x_{min}$ is the minimum cell size and $0<\beta\le 1$ is the CFL number.

To demonstrate Point (ii), we first rewrite Eq. \eqref{eq:face-phi} in terms of the original distribution function as
\begin{eqnarray}
\label{eq:AP-face}
\phi_{\alpha}^{n+1/2}(\x_b)&=&
A(\tau, \Delta t)\left[{\phi}_{\alpha}^{n}(\x_c)+(\x'_{\alpha}-\x_c) \cdot\bm{\delta}_c{\phi}_{\alpha}^{n}\right]\nonumber \\
&&+
B(\tau, \Delta t)\left[{\phi}_{\alpha}^{eq,n}(\x_c)+(\x'_{\alpha}-\x_c) \cdot\bm{\delta}_c{\phi}_{\alpha}^{eq,n}\right]
+
B(\tau, \Delta t)\phi_{\alpha}^{eq,n+1/2}(\x_b).
\end{eqnarray}
where the two parameters $A$ and $B$ are
$$
A(\tau, \Delta t)=\dfrac{4\tau-\Delta t} {4\tau+\Delta t},\quad B(\tau,\Delta t)=\dfrac{\Delta t} {4\tau+\Delta t}.
$$
Note that in the DUGKS the discrete distribution functions are assumed to be piecewise linear in each cell, and therefore Eq. \eqref{eq:AP-face} can also be expressed as
\begin{eqnarray}
\label{eq:AP-facex}
\phi_{\alpha}^{n+1/2}(\x_b)&=&
A(\tau, \Delta t)\left[{\phi}_{\alpha}^{n,c}(\x_b)-\tfrac{\Delta t}{2}\v_{\alpha} \cdot\bm{\delta}_c{\phi}_{\alpha}^{n}\right]\nonumber \\
&&+
B(\tau, \Delta t)\left[{\phi}_{\alpha}^{eq,n}(\x_b)-\tfrac{\Delta t}{2}\v_{\alpha} \cdot\bm{\delta}_c{\phi}_{\alpha}^{eq,n}\right]
+
B(\tau, \Delta t)\phi_{\alpha}^{eq,n+1/2}(\x_b),
\end{eqnarray}
where $\phi_{\alpha}^{n,c}(\x_b)=\phi_{\alpha}^{n}(\x_c)+(\x_b-\x_c)\cdot\delta_c\phi_{\alpha}^{n}$ is the value of the distribution function at cell interface $\x_b$ determined from cell $\x_c$.

The AP property of the DUGKS depends on the parameters $A$ and $B$. If we define the macroscopic diffusive length and time scales, $l_0$ and $t_0$, respectively, such that $\lambda_0/l_0=\epsilon \sim \mbox{Kn}$ and $\tau_0/t_0=\epsilon^2$, the parameters $A$ and $B$ can the be expressed as
\begin{equation}
\label{eq:AB}
A(\tau, \Delta t)=\dfrac{4\epsilon^2\tau'-\Delta t'} {4\epsilon^2\tau'+\Delta t'},\quad B(\tau,\Delta t)=\dfrac{\Delta t'} {4\epsilon^2\tau'+\Delta t},
\end{equation}
where $\tau'=\tau/\tau_0$ and $\Delta t'=\Delta t/t_0$ are the dimensionless relaxation time and time step, respectively.
In the ballistic limit ($\epsilon \to \infty$), we can obtain that $A=1$ and $B=0$, and thus
\begin{equation*}
\phi_{\alpha}(\x_b)={\phi}_{\alpha}^{n,c}(\x_b)-\tfrac{\Delta t}{2}\v_{\alpha} \cdot\bm{\delta}_c{\phi}_{\alpha}^{n}=\phi_{\alpha}(\x_b-\v_{\alpha}\Delta t/2),
\end{equation*}
which is just a solution of the free transport kinetic equation. Specifically, in the 1D case as sketched in Fig. \ref{fig:Cell}, the reconstructed interface distribution function is
\begin{eqnarray}
\phi_{\alpha,j+1/2}^{n+1/2}&=& \left[{\phi}_{\alpha,j+1/2}^{n,L}-\tfrac{1}{2}v_{\alpha}\Delta t \delta_j{\phi}_{\alpha}^{n}\right]H(v_\alpha)
+\left[{\phi}_{\alpha,j+1/2}^{n,R}-\tfrac{1}{2}v_{\alpha}\Delta t \delta_{j+1}{\phi}_{\alpha}^{n}\right]\bar{H}(v_\alpha)
\end{eqnarray}
where $\phi_{j+1/2}^L$ and $\phi_{j+1/2}^R$ are the left and right values of the distribution functions at cell interface $j+1/2$, respectively, $H$ is the Heaviside function, $H(x)=1$ if $x>0$, and 0 otherwise, while $\bar{H}=1-H$. Therefore, the DUGKS Eq. \eqref{eq:center-phi} can be expressed explicitly as (note that $\tilde{\phi}_{\alpha}=\tilde{\phi}_{\alpha}^+ = \phi_{\alpha}$ in this limit)
\begin{eqnarray}
\dfrac{\phi_{\alpha, j}^{n+1}-\phi_{\alpha, j}^{n}}{\Delta t}
&+&\dfrac{v_{\alpha}}{\Delta x_j}
\left\{
H(v_{\alpha})\left[\phi_{\alpha,j+1/2}^{n,L}-\phi_{\alpha,j-1/2}^{n,L}\right]+\bar{H}(v_{\alpha})\left[\phi_{\alpha,j+1/2}^{n,R}-\phi_{\alpha,j+1/2}^{n,R}\right]
\right\}\nonumber \\
&-&\dfrac{v_{\alpha}^2\Delta t}{2\Delta x_j}
\left\{
H(v_{\alpha})\left[\delta_j\phi_{\alpha}^{n}-\delta_{j-1}\phi_{\alpha}^{n}\right]
+\bar{H}(v_{\alpha})\left[\delta_{j+1}\phi_{\alpha}^{n}-\delta_{j}\phi_{\alpha}^{n}\right]
\right\}
=0,
\end{eqnarray}
which is a consistent finite-volume scheme of Lax-Wendroff type for the kinetic equation without scattering. This result suggests that the DUGKS is AP in the ballistic limit.

Now we discuss the AP property of the DUGKS in the diffusive limit ($\epsilon\to 0$). Under this limit, the distribution function can be approximated as (see Appendix A),
\begin{equation}
\label{eq:1st-order expansion}
\phi_{\alpha} = \phi_{\alpha}^{eq}-\tau\v_{\alpha}\cdot\nabla\phi \approx \phi_{\alpha}^{eq}-\tau\v_{\alpha}\cdot\bm{\delta}_c \phi.
\end{equation}
Substituting this approximation into Eq. \eqref{eq:AP-facex} we can obtain that
\begin{eqnarray}
\label{eq:AP-facey}
\phi_{\alpha}^{n+1/2}(\x_b)&=&
(1-B)\phi_{\alpha}^{eq,n,c}(\x_b)-\tau\v_{\alpha}\cdot\bm{\delta}_c\phi_{\alpha}^{eq,n,c} + B \phi_{\alpha}^{eq,n+1/2}(\x_b) + O(\tau\Delta t)\nonumber\\
&=&(1-B)\phi_{\alpha}^{eq,n}(\x_b)-\tau\v_{\alpha}\cdot\bm{\delta}\phi_{\alpha}^{eq,n} + B \phi_{\alpha}^{eq,n+1/2}(\x_b),
\end{eqnarray}
where we have assumed that in diffusive limit the distribution function is smooth across cell interfaces, i.e., $\phi_{\alpha}^{eq,n,c}=\phi_{\alpha}^{eq,n}$ and $\bm{\delta}_c\phi=\bm{\delta}\phi$. On the other hand, based on the properties of the angular quadrature given by Eq. \eqref{eq:w}, we have
\begin{equation}
\sum_{\alpha}{w_{\alpha} \v_{\alpha}\phi_{\alpha}^{eq}}=\bm{0}, \quad \sum_{\alpha}{w_{\alpha} \v_{\alpha}\v_{\alpha}\phi_{\alpha}^{eq}}=\dfrac{v_g^2}{3}E\bm{I}.
\end{equation}
Therefore, the macroscopic flux across the cell interface $\x_b$ can be obtained,
\begin{equation}
\mathcal{F}(\x_b)=\sum_{\alpha}{w_{\alpha}\v_{\alpha}\phi_{\alpha}^{n+1/2}(\x_b)}=-\dfrac{1}{3}\tau v_g^2 \bm{\delta} E^n(\x_b)=-\kappa \bm{\delta} E^n(\x_b).
\end{equation}
Then by taking the zeroth angular moment of Eq. \eqref{eq:center-phi} we can obtain that
\begin{equation}
\dfrac{E_j^{n+1}-E_j^{n}}{\Delta t} - \dfrac{1}{|V_j|}\sum_{\x_b\in {\cal N}_j}{ \kappa \bm{\delta} E^n(\x_b) } = 0.
\end{equation}
Specifically, for the 1D case, the above equation becomes
\begin{equation}
\dfrac{E_j^{n+1}-E_j^{n}}{\Delta t} - \dfrac{1}{\Delta x_j}{\kappa [\delta E_{j+1/2}^n -\delta E_{j-1/2}^n}] = 0,
\end{equation}
which is just an explicit solver for the diffusion equation \eqref{eq:MacroEq}, suggesting that the DUGKS is also AP in the diffusive limit.

\section{Boundary conditions}
Generally three types of boundary conditions are used to describe the interactions between phonons and material boundaries, \cite{ref:Narumachi,ref:Struch} namely, specular reflection, thermalization, and diffusive reflection.
The specular reflection is similar to that in classical gas kinetic theory. It assumes that a phonon is
just reflected back to the domain with a reflection angle equal and opposite to the incident one after it hits the surface, such that the phonon energy density for directions $\s_{\alpha}$ entering the domain is given by
\begin{equation}
\phi_{\alpha}=\phi_{\alpha'}, \quad \s_\alpha\bm{n} > 0,
\end{equation}
where $\s_{\alpha'}=\s_{\alpha}-2(\s_{\alpha}\cdot\bm{n})\bm{n}$, with $\bm{n}$ the outward unit normal vector to the wall pointing into the domain. Therefore, for the specular reflection, the phonon energy is conserved and there is no energy transfer across the boundary.

In thermalization boundary condition,  a phonon is absorbed as it strikes the boundary, and a new phonon in thermal equilibrium with boundary temperature is emitted into the domain. Therefore, the reflected phonon can be expressed as
\begin{equation}
\phi_{\alpha}=\phi_{\alpha}^{eq}(T_s),
\end{equation}
where $T_s$ is the temperature at the surface. Since $\tilde{\phi}_{\alpha}$ and $\bar{\phi}_{\alpha}$ are linear combinations of $\phi_{\alpha}$ and $\phi_{\alpha}^{eq}$, the thermalization boundary condition in the DUGKS can also be expressed as
\begin{equation}
\tilde{\phi}_{\alpha}=\tilde{\phi}_{\alpha}^{eq}(T_s) , \quad \bar{\phi}_{\alpha}=\bar{\phi}_{\alpha}^{eq}(T_s).
\end{equation}
It can be seen that the thermalization boundary condition for phonon transport is very similar to the Maxwell diffuse scattering in the classical gas kinetic theory, which assumes the distribution function of reflected particles follows a
Maxwellian one with the wall temperature and velocity. It is obvious that thermalization boundary condition allows for energy transfer across the surface.

In the diffusive reflection boundary condition, which should not be confused with the Maxwell diffuse scattering in gas kinetic theory, the phonons hitting the surface are reflected with equal probability along all possible angles, namely,
\begin{equation}
f(\s)=\left[\int_{\bm{s}'\cdot\bm{n}<0}{ (\bm{s}'\cdot\bm{n})\,d\Omega}\right]^{-1} \int_{\bm{s}'\cdot\bm{n}<0} { (\bm{s}'\cdot\bm{n}) f(\s')\,d\Omega },
\end{equation}
or in terms of the discrete phonon energy distribution function,
\begin{equation}
\phi_{\alpha}=\left[\sum_{\s_{\beta}\cdot\bm{n} <0}{w_{\beta}(\s_{\beta}\cdot\bm{n})}\right]^{-1}\sum_{\s_{\beta}\cdot\bm{n} <0}{w_{\beta}(\s_{\beta}\cdot\bm{n})\phi_{\beta}}.
\end{equation}
Since the phonons hitting the surface are all reflected back to the domain, the total energy in the diffusive reflection boundary condition is conserved and no heat transfer occurs across the boundary.

\section{Numerical tests}
\begin{figure}
\includegraphics[width=0.45\textwidth]{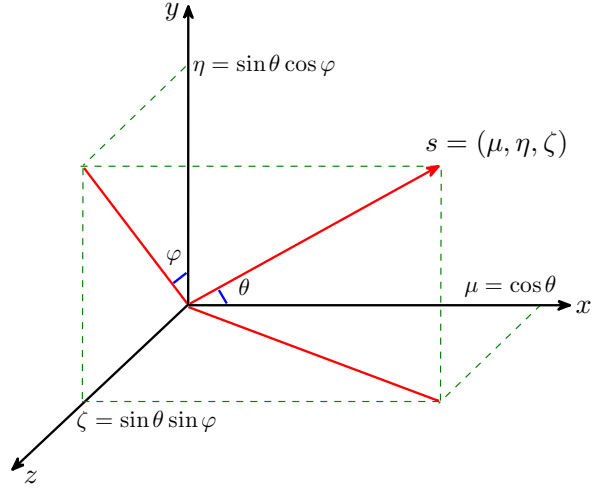}
\caption{Local coordinate system.}
\label{fig:Coordinate}
\end{figure}

In this section we will apply the DUGKS to several heat transfer problems with different Knudsen numbers to test its performance. In the simulations, the CFL number is fixed at 0.9 unless stated otherwise. The van Leer limiter is employed to determine the slope $\bm{\delta}_c\phi_{j,\alpha}^+$ in each cell for problems as described in Subsec. \ref{subsec:Flux}. The local coordinate system used in the simulations is shown in Fig. \ref{fig:Coordinate}, where $\theta\in[0, \pi]$ and $\phi\in [0, 2\pi]$ are the polar and azimuthal angles.

\subsection{Heat Conduction Across a Film}
First we consider a dielectric film of thickness $L$, the temperatures of the two boundaries located at $x=0$ and $x_L$ maintain at $T_0$ and $T_L$, respectively. The problem is 1D and the energy distribution function $e''$ depends on spacial variable $x$ and angular variable $s_x=\mu=\cos\theta$ only, such that we can define a reduced distribution function $e''(x,\mu)$ by integrating $e''(\s)$ with respect to $\varphi$,
\begin{equation}
e''(x,\mu)=\int_0^{2\pi}{e''(x,\mu,\sin\theta\cos\varphi,\sin\theta\sin\varphi)\, d\varphi },
\end{equation}
and the BTE for the phonon energy density at steady state becomes
\begin{equation}
\mu\dpart{e''}{x}=-\dfrac{1}{\lambda}\left[e''-e^{eq}\right],
\end{equation}
where $e^{eq}(x)=E(x)/2$ is the reduced equilibrium distribution for energy density. With the thermalizing boundary conditions at the two surfaces, the solution of this problem can be expressed as, \cite{ref:ChenBook}
\begin{equation}
\label{eq:1DE}
E^*(x^*)=\dfrac{1}{2}\left[E_2(x^*)+\int_0^{\xi} {  E^*(x') E_1(|x^*-x'|)\, dx'}   \right],
\end{equation}
where $x^*=x/\lambda$ is the nondimensional position, $E^*=(E-E_0)/(E_L-E_0)$ is the nondimensional energy with $E_0=C T_0$ and $E_L=C T_L$, $\xi=L/\lambda=1/\mbox{Kn}$ is the acoustic thickness, and $E_n(x)=\int_0^1{t^{n-2}\exp(-x/t)\, dt}$ is the exponential integral function. The dimensionless heat flux can be expressed as
\begin{equation}
\label{eq:1Dq}
q^*=\dfrac{1}{v_g(E_0-E_L)}\int_0^{L} {v_g\mu e''(x,\mu)\, d\mu}=1-2\int_0^{\xi} {  E^*(x') E_2(x')\, dx'},
\end{equation}
which is a constant across the domain. The two integration equations \eqref{eq:1DE} and \eqref{eq:1Dq} can be solved numerically to give an ``numerical exact" solutions. \cite{ref:ChenBook} Here we use 4000 points such that the solutions are mesh independent.
\begin{figure}
\includegraphics[width=0.45\textwidth]{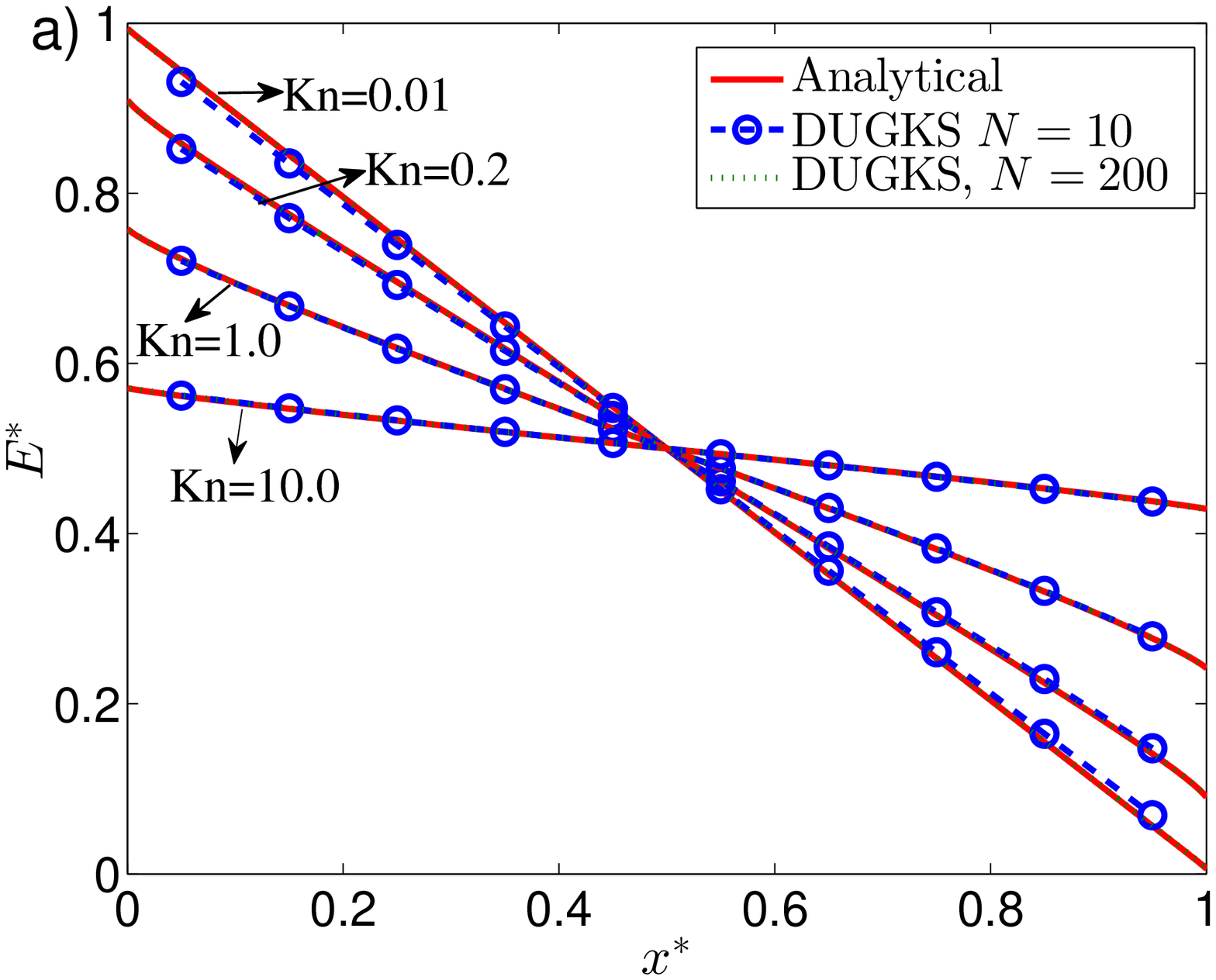}
\includegraphics[width=0.45\textwidth]{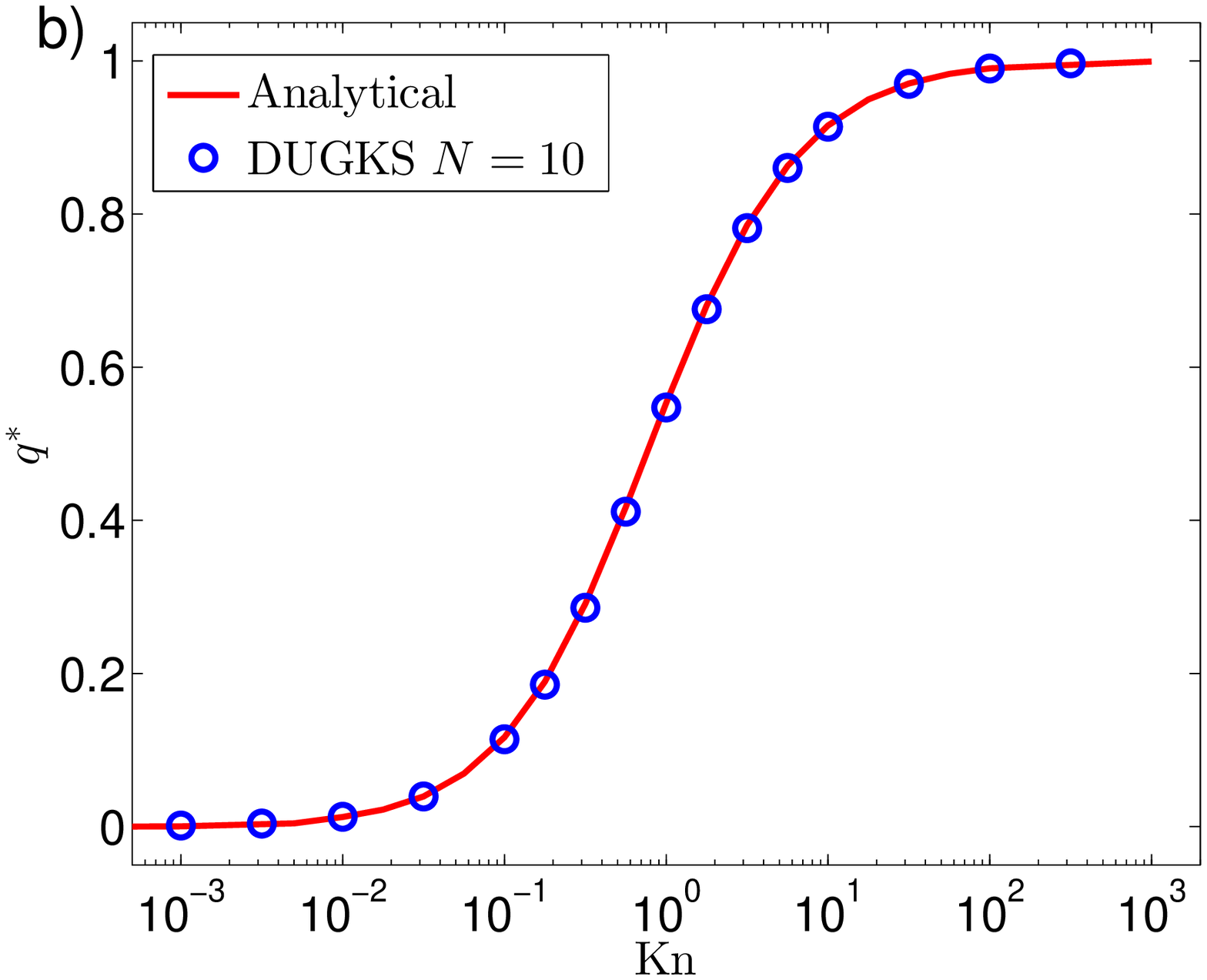}
\caption{Dimensionless energy and heat flux across a film.}
\label{fig:Film}
\end{figure}

The DUGKS is applied to this 1D problem at various Knudsen numbers, and in the simulations the Gauss-Legendre quadrature with $N_\mu=100$ points is employed to evaluate the angular moments with respect to the directional cosine $\mu$, i.e,
\begin{equation}
E(x)=\int_{-1}^1{e''(\x,\mu)\, d\mu}=\sum_{\alpha=1}^{N_\mu}{w_\alpha \phi_{\alpha}(\x)}, \quad q(x) =\int_{-1}^1{v_g\mu e''(\x)\, d\mu} = \sum_{\alpha=1}^{N_\mu}{w_\alpha v_g\mu_{\alpha}\phi_{\alpha}(\x)}.
\end{equation}
It should be noted that here we choose $N_\mu=100$ such that the moments can be evaluated accurately for all of the Knudsen numbers considered. Actually, a quadrature with much less angular points can be employed as $\mbox{Kn}\le 1$, say $N_\mu=16$.

The dimensionless energy $E^*$  heat flux $q^*$ with mesh resolutions of $N=10$ and 200  are shown in Fig. \ref{fig:Film}. It can be observed that the DUGKS results agree well with the analytical solutions even with the coarse mesh, suggesting that the present scheme exhibits low numerical diffusion and is insensitive to mesh resolutions. The results also show the uniform stable property of the present DUGKS in the sense that the time step is solely determined by the CFL condition and is independent of the relaxation time. In other words, the time step $\Delta t$ (or cell size $\Delta x$) is not required to be smaller than the relaxation time $\tau$ (or mean free path $\lambda$). Actually, with the coarse mesh ($N=10$),  the value of $\Delta x/\lambda =1/(N\mbox{Kn})$ ranges from 0.001 to 100 as $\mbox{Kn}$ changes from $10^{-3}$ to $10^{2}$, and $\Delta t/\tau$ ranges from $9\times 10^{-4}$ to 90.

\begin{figure}
\includegraphics[width=0.45\textwidth]{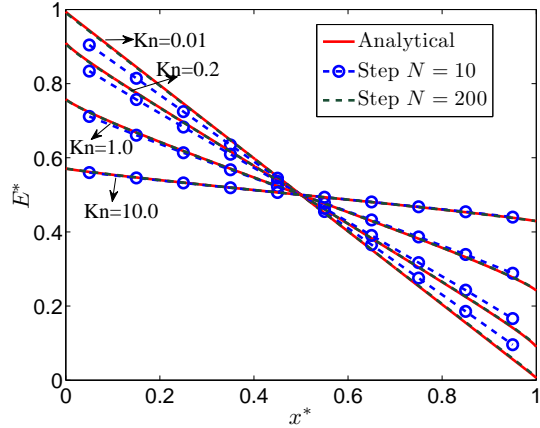}
\caption{Dimensionless energy from the step method across a film.}
\label{fig:Step}
\end{figure}

For comparison, we also applied the implicit finite-difference method with step (upwind) discretization of the convection term of Eq. \eqref{eq:BGK-phi}, which is widely used for solving the phonon BTE, \cite{ref:Ye2015,ref:Narumachi,ref:YangChen2005} to the present problem. The profiles of the dimensionless energy $E^*$ at various Kn are shown in Fig. \ref{fig:Step}. It can be observed with the fine mesh of $N=200$, the results agree well with the analytical solutions as $\mbox{Kn}$ changes from 0.01 to 10. This is reasonable since $\Delta x/\lambda\le 0.5$ in all of the considered cases, and the numerical diffusion, which is proportional to $\Delta x$, is less than the physical one with this mesh resolution. On the other hand, with the coarse mesh, clear deviations from the analytical solutions can be observed as $\mbox{Kn}=0.01$ and 0.2.

\begin{figure}
\includegraphics[width=0.6\textwidth]{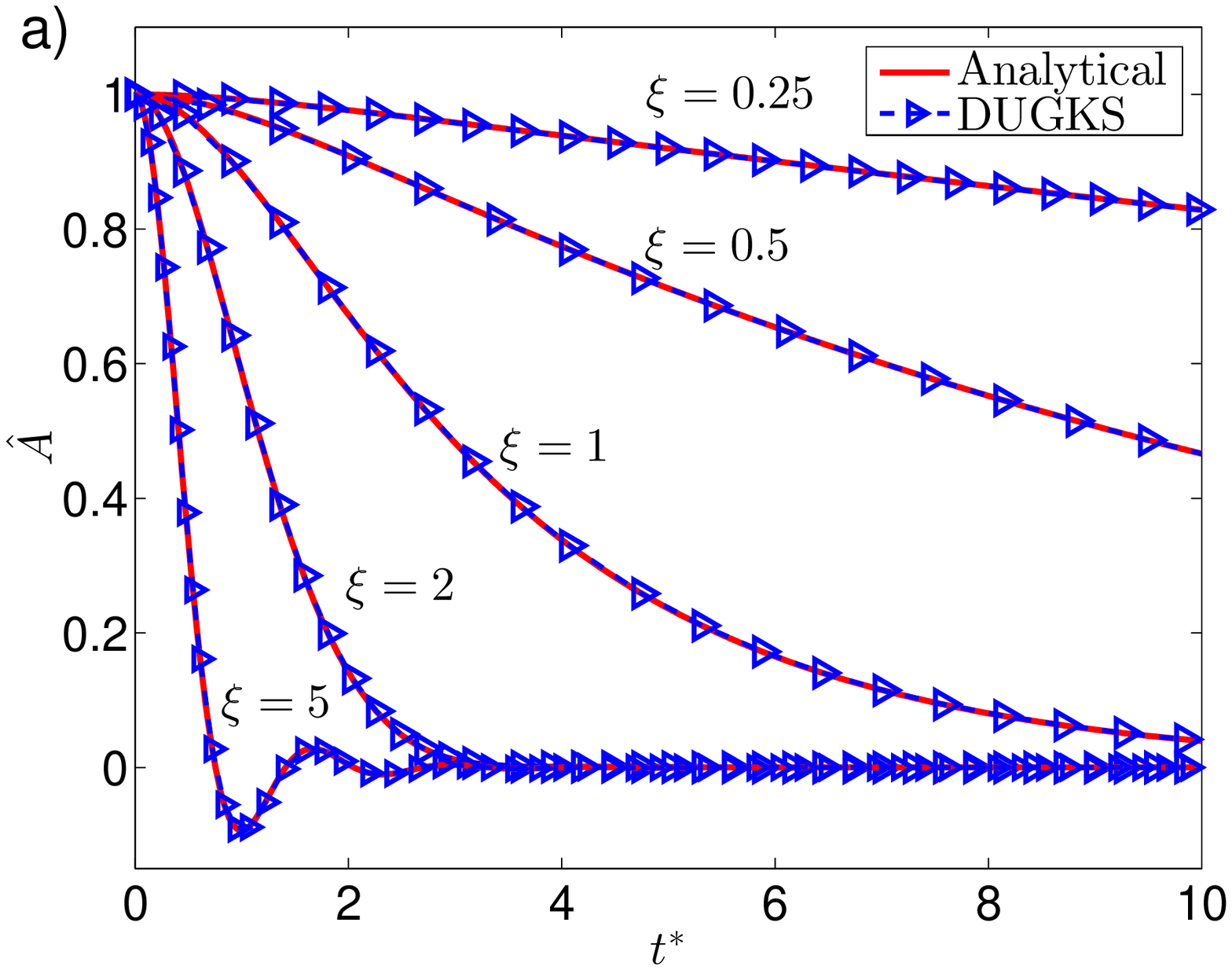}\vfill
\includegraphics[width=0.45\textwidth]{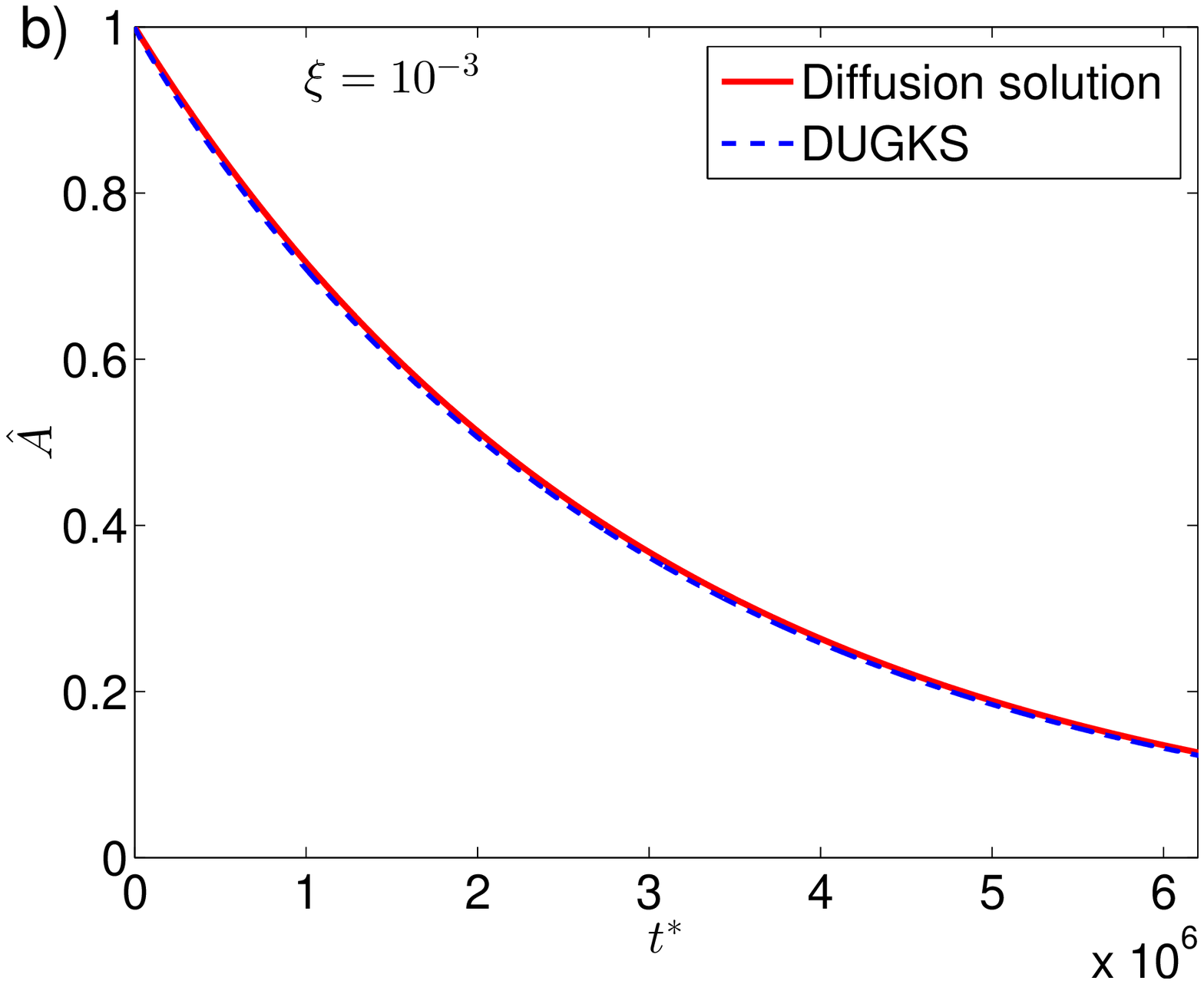}\hfill
\includegraphics[width=0.45\textwidth]{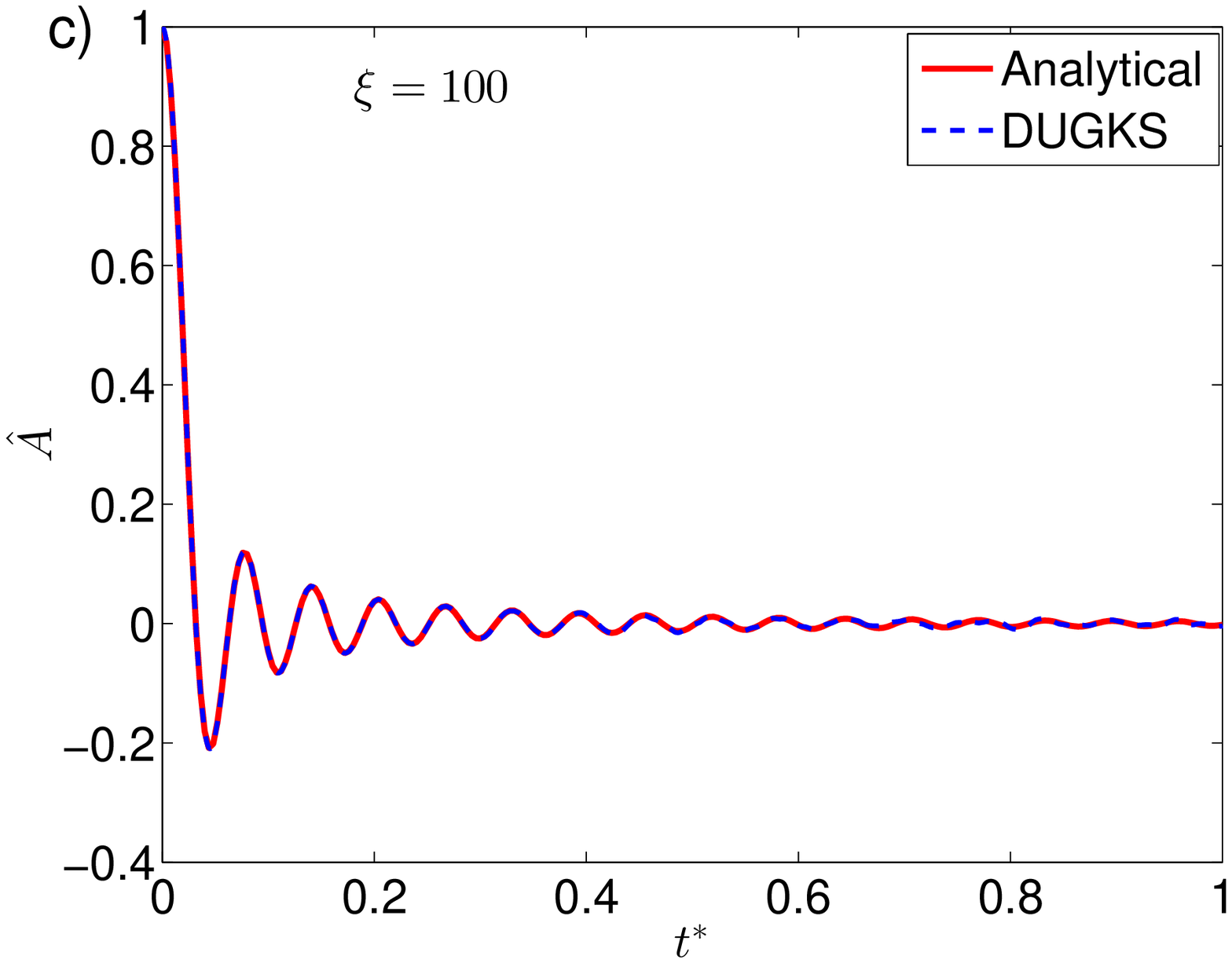}
\caption{Decaying of the dimensionless amplitude of temperature variation $\hat{A}=A/A_0$ against dimensionless time $t^*=t/\tau$ at different Knudsen numbers $\xi=2\pi\mbox{Kn}$. (a) Finite values of $\xi$; (b) Diffusive limit; (c) Ballistic limit. }
\label{fig:TTG-T}
\end{figure}

\subsection{Heat relaxation of transient thermal grating}
The transient thermal grating (TTG) is a technique for measuring thermal conductivity and phonon mean free path of a material. \cite{ref:TTG_Rogers,ref:TTG_Minnich2012,ref:TTG_Collins2013,ref:TTG_Hua2014} Here we considered the thermal relaxation process of a 1D TTG, \citep{ref:TTG_Collins2013} where initially two crossed laser pulses are imposed to produce a spatially sinusoidal temperature variation in the medium,
\begin{equation}
T(x,0)= T_b + A_0\cos(\theta x),
\end{equation}
where $T_b$ is the background temperature, $A_0$ is the amplitude of the temperature variation, and $\theta =2\pi/l$ is the wave number with $l$ being the grating period. As the strength of the pulse is weak, the BTE can be linearized, and the temperature deviation from the background temperature, $\Delta T=T-T_b$, can be approximated as $\Delta T(x,t)=A(t)\cos(\theta x)$, where the amplitude $A$ can be obtained analytically, \cite{ref:TTG_Collins2013}
\begin{equation}
\label{eq:TTG}
\hat{A}(t^*)=\sinc (\xi t^*)e^{-t^*} + \int_0^{t^*}{\hat{A}(t')\sinc [\xi(t'-t^*)]e^{(t'-t^*)}\, dt'},
\end{equation}
with $\hat{A}=A/A_0$ and $t^*=t/\tau$, and $\xi=2\pi\mbox{Kn}$ is the rarefaction parameter with the Knudsen number defined as $\mbox{Kn} =v_g\tau/l$. Equation \eqref{eq:TTG} is a second kind Volterra integral equation and can be solved using standard numerical techniques. \cite{ref:Vol} In the diffusive limit ($\xi\to 0$), it can be shown that
\begin{equation}
\hat{A}(t^*)=e^{-\gamma t^*},
\end{equation}
where $\gamma=\xi^2/3$; while in the ballistic limit ($\xi\to\infty$), we have $\hat{A}(t^*)=\sinc(\xi t^*)$, which shows strong oscillations.

We simulate the thermal relaxation process of the 1D TTG at different Knudsen numbers. The length of the computation domain is taken to be $L=2 l$, and a uniform grid of 100 points (i.e., 50 points in one grating period) is used for all cases. The directional cosine space $-1\le \mu\le 1$ is discretized using the  Gauss-Legendre quadrature with $N_\mu=100$ points. Periodic boundary conditions are applied to the left ($x=0$) and right ($x=L$) boundaries. In Fig. \ref{fig:TTG-T} the time histories of the amplitude of the temperature variation are shown for different values of $\xi$. It can be seen that the numerical results of the DUGKS are in excellent agreement with the analytical solutions for different Knudsen numbers, ranging from diffusive limit to ballistic limit. The results again confirm the asymptotic preserving properties of the present DUGKS method for modeling multiscale heat transfer.

\subsection{Heat transfer in a 2D square domain}
In the above subsections the DUGKS was tested by 1D steady and 1D unsteady heat transfer problems. We now consider a two-dimensional problem in a square medium of length $L$. Initially the temperature of the medium is set to be a uniform $T_0$, and then the temperature at the bottom side ($y=0$) is raised to and maintained at $T_1>T_0$. Thermalization boundary conditions are assumed on the four sides, and the Knudsen number of the system is defined as $\mbox{Kn}=\lambda/L$. This problem was studied recently by solving the BTE with a DOM coupled with finite-element scheme, \cite{ref:DOM-FE} and a similar radiative heat transfer was studied early, \cite{ref:Crosbie1984} which is actually identical to the phonon transfer as the medium is exposed to isotropically scattering.

Simulations at different Knudsen numbers are conducted based on a uniform mesh $N_x\times N_y=60\times 60$ in physical space, and the directional cosine $\mu\in[-1, 1]$ (i.e., $0\le \theta\le\pi$) is discretized with the 32-point Gauss-Legendre quadrature, while the azimuthal angular space $\varphi\in[0, \pi]$ (not $[0, 2\pi]$ due to symmetry) is discretized with the 16-point Gauss-Legendre quadrature. Our simulations show that the meshes are sufficient to obtain convergent results.

\begin{figure}
\includegraphics[width=0.6\textwidth]{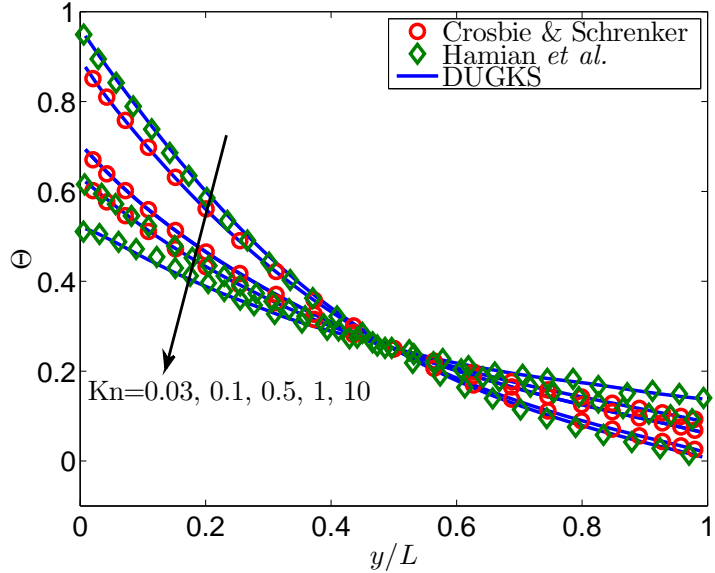}\vfill
\caption{Normalized temperature profile along the vertical centerline at different Knudsen numbers. Reference data are taken from Refs. [\onlinecite{ref:DOM-FE}] and [\onlinecite{ref:Crosbie1984}]}
\label{fig:2D}
\end{figure}

In Fig. \ref{fig:2D} the normalized temperatures, $\Theta=(T-T_0)/(T_1-T_0)$, are shown along the vertical centerline at $x=L/2$ as the system reaches the steady state. For comparison, we also present the solutions of DOM for phonon BTE \cite{ref:DOM-FE} and the numerical analytical solutions of the integral equation
describing radiative transfer in a 2D isotropically scattering medium. \cite{ref:Crosbie1984} It can be seen that the results of the DUGKS agree quite well with the reference data for all Knudsen numbers considered here. It is noted that some slight oscillations appear in the temperature profile predicted by the DOM at $\mbox{Kn}=10$. \cite{ref:DOM-FE} This is caused by the ray effect due to the insufficient discretization in angular space, which was based on a $N_{\mu}\times N_{\varphi}=32\times 8$ Gaussian quadrature. \cite{ref:DOM-FE} On the other hand, the temperature profile from the present DUGKS does not show this behavior with the $32\times 16$ angular discretization.

\begin{figure}
\includegraphics[width=0.45\textwidth]{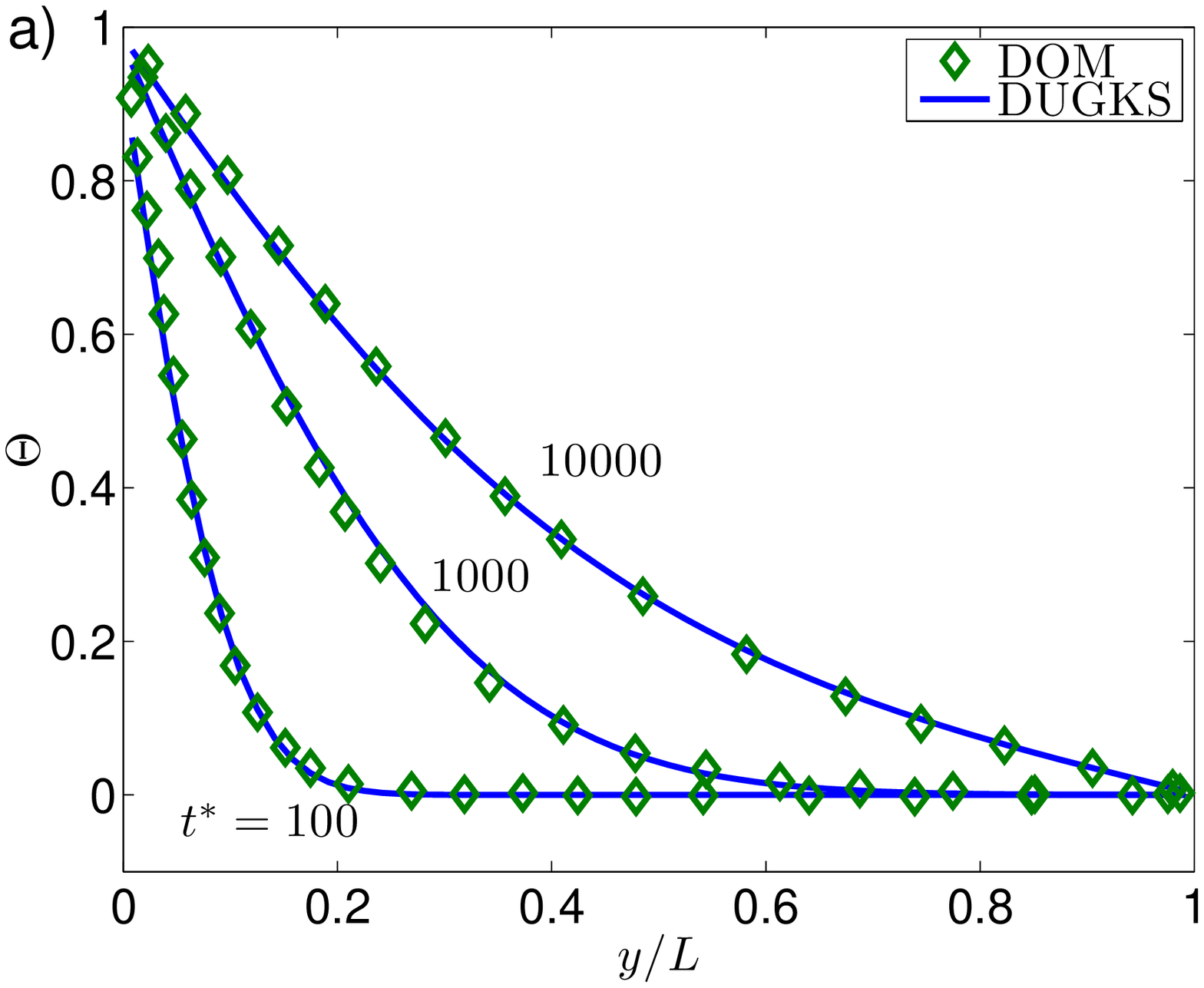}\hfill
\includegraphics[width=0.45\textwidth]{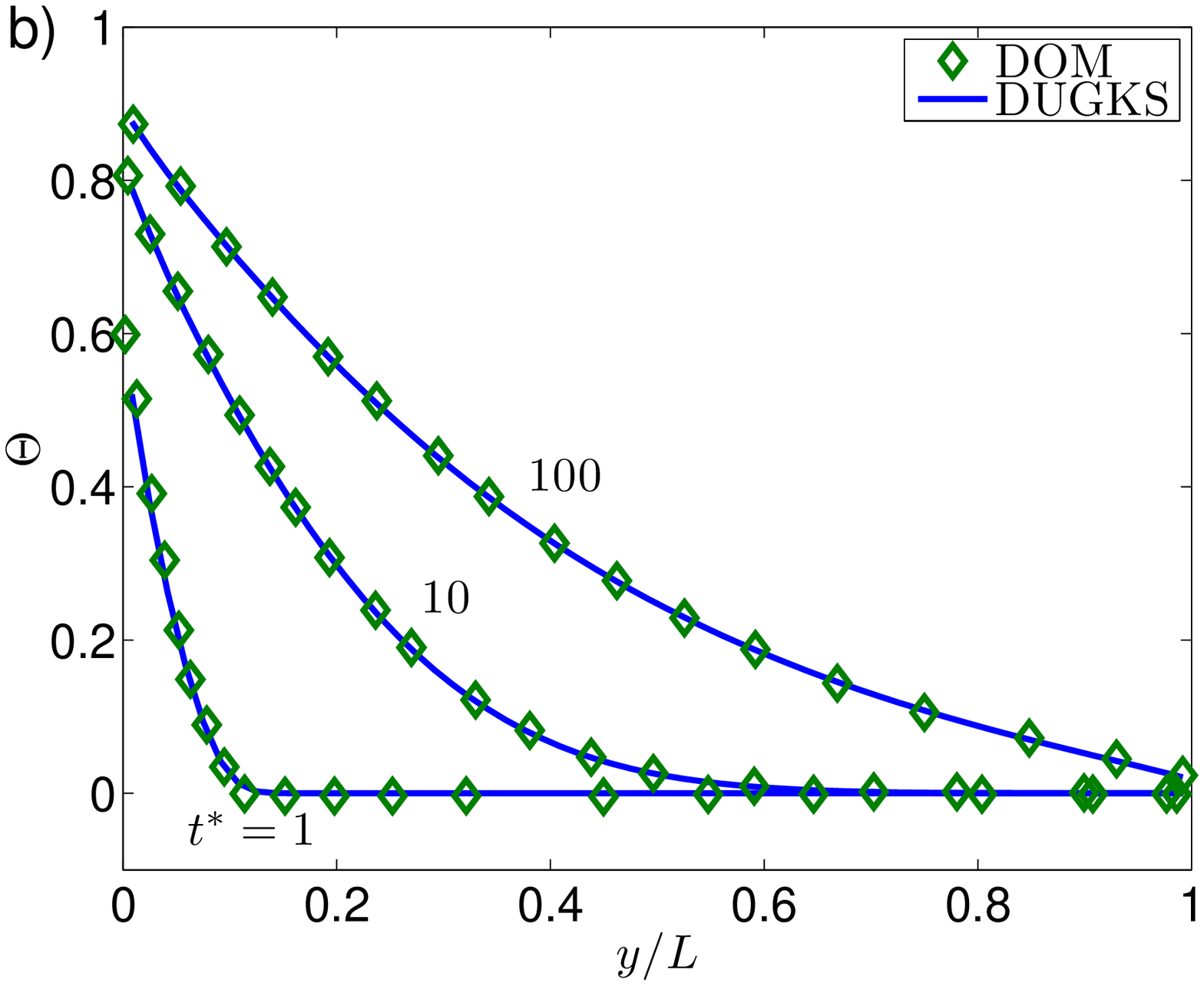}\vfill
\includegraphics[width=0.45\textwidth]{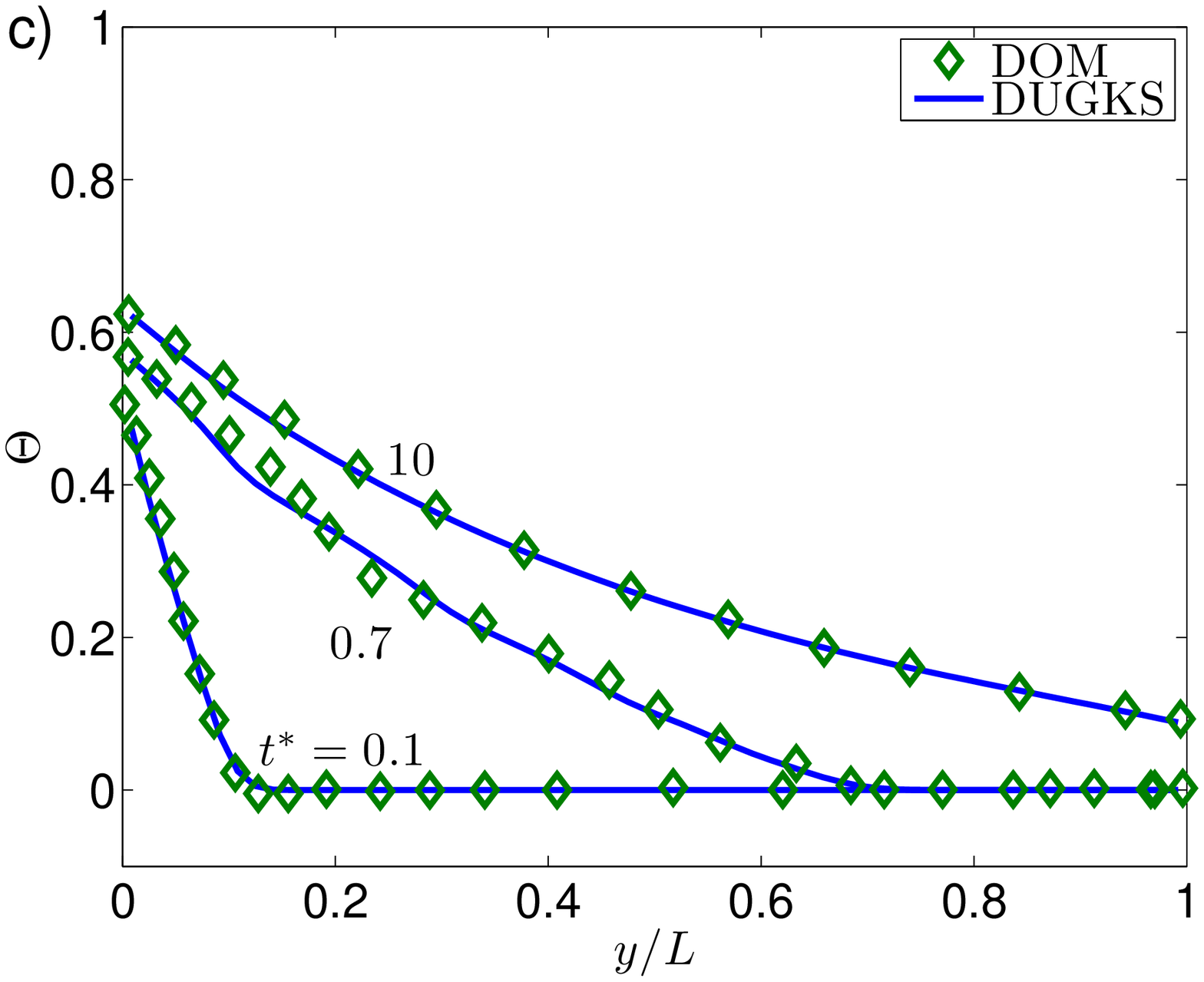}\hfill
\includegraphics[width=0.45\textwidth]{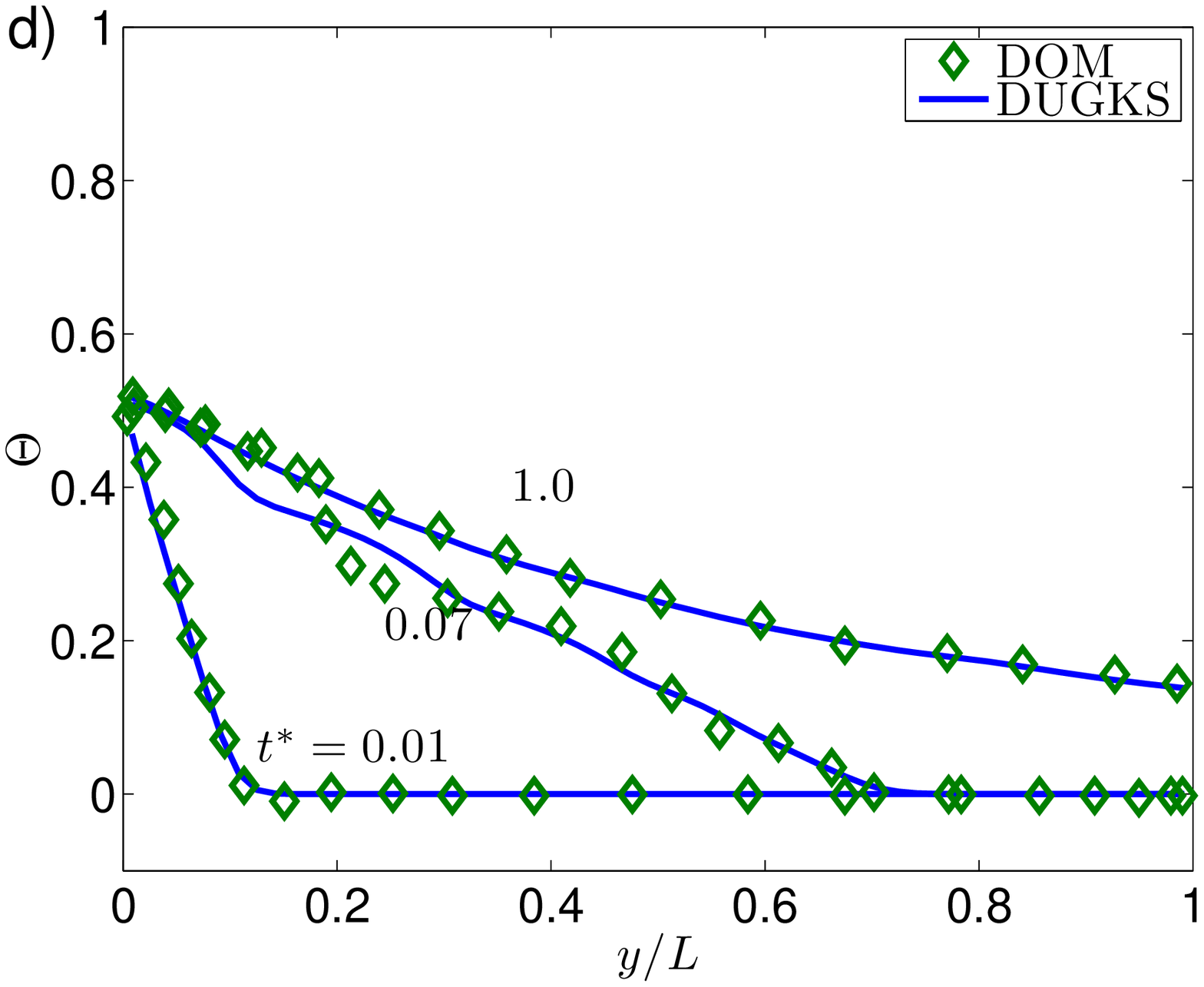}
\caption{Transient temperature profile along the vertical centerline at different Knudsen numbers. The DOM data are taken from Ref. [\onlinecite{ref:DOM-FE}].}
\label{fig:time2D-T}
\end{figure}

The transient behavior of this 2D problem is also compared with the results of the DOM. \cite{ref:DOM-FE} The temperature profiles along the vertical centerline of the medium are shown in Fig. \ref{fig:time2D-T} at different times of $t^*=t/\tau$ for different Knudsen numbers ranging from diffusive to ballistic regimes. It can be observed that the DUGKS results agree qualitatively well with the DOM predictions from initial to later times in all cases considered. Specifically, as $\mbox{Kn}=0.01$, the heat transfer is diffusive, and no obvious temperature jump occurs at the top and bottom boundaries at different times. As Kn increases to 0.1, i.e., the problem falls in the near diffusive regime, and temperature jump appears on both boundaries, particularly on the bottom hot surface, where the jump decreases with time. At $\mbox{Kn}=1$, temperature jump is more significant, but transient jump becomes smaller than that for the case of $\mbox{Kn}=0.1$. As $\mbox{Kn}=10$, the heat transfer is dominated by ballistic effect, and the temperature jump maintains nearly constant on the hot surface, although transient changes can still be observed on the bottom cold surface. It is also noted that some differences between the present results and the DOM data for the case of $\mbox{Kn}=10$ at $t^*=0.07$, which can again be attributed to the insufficient angular discretization
of the DOM. Overall, the results of this 2D heat transfer problem confirm the AP and uniform stable properties of the proposed DUGKS for simulating heat transport process from diffusive to ballistic regimes.

\section{summary}

In this work we develop a discrete unified gas kinetic scheme for multiscale heat transfer based on the phonon Boltzmann transport equation. With the coupled treatment of the scattering and transport of phonons, the scheme is asymptotic preserving, and the limitation on the time step being less than the relaxation time in diffusive regime is removed. The method has been validated in two 1D and a 2D problems at different Knudsen numbers, and the results are in excellent agreement with analytical or benchmark data.

In the present work, the proposed DUGKS is designed based on the gray BTE model, where the frequency dependence of the distribution function, relaxation time, and group velocity, are ignored. However, the current methodology can be extended to non-gray BTE models, which will be studied in our subsequent works.

\acknowledgments{ZLG acknowledges the support by the National
Natural Science Foundation of China (51125024), and part of the
work was carried out during his visit to the Hong Kong University
of Science and Technology. The research of KX was supported by Hong Kong Research Grant Council
(620813, 16211014, 16207715) and HKUST research fund (PROVOST13SC01, IRS15SC29, SBI14SC11).
}
\appendix
\section{Analysis of the BTE with diffusive scaling}
In order to analyze the limit behavior of the BTE in the diffusive regime, we here make use of the diffusive scaling analysis introduced by Sone in gas kinetic theory, \cite{ref:Sone} which was also used in the analysis of the Unified Gask Kinetic Scheme for radiative transper problems. \cite{ref:Luc} With the diffusive scaling, i.e., $t_0=t/\epsilon^2$ and $x_0=x/\epsilon$ with $\epsilon\sim\mbox{Kn}$ being a small parameter, we have
\begin{equation}
\partial_{t}=\epsilon^2\partial_{t_0}, \quad \partial_{x_i} =\epsilon \partial_{x_0}.
\end{equation}
Then the BTE \eqref{eq:Boltzmann-BGK} can be rewritten as
\begin{equation}
\label{eq:A-BGK}
\epsilon^2\partial_{t_0} {e''}+\epsilon \v_{\alpha}\cdot\nabla_0 e''=-\dfrac{1}{\tau}\left[e''-e^{eq}\right].
\end{equation}
We further expand $e''$ in a power series of $\epsilon$,
\begin{equation}
\label{eq:Ae-expansion}
e''=e^{(0)}+\epsilon e^{(1)} + \epsilon^2 e^{(2)} + \cdots.
\end{equation}
Substituting this expansion into Eq. \eqref{eq:A-BGK}, we can obtain the equations in the consecutive orders of $\epsilon$,
\begin{subequations}
\label{eq:Aorder}
\begin{align}
\epsilon^0:&\qquad e^{(0)}=e^{eq},\label{eq:Aorder-0} \\
\epsilon^1:&\qquad \v\cdot\nabla_0 e^{(0)}=-\dfrac{1}{\tau} e^{(1)},\label{eq:Aorder-1}\\
\epsilon^2:&\qquad \partial_{t_0} e^{(0)}+\v\cdot\nabla_0 e^{(1)}=-\dfrac{1}{\tau} e^{(2)}. \label{eq:Aorder-2}
\end{align}
\end{subequations}
From Eqs. \eqref{eq:Ae-expansion} and \eqref{eq:Aorder-0}, and recalling the energy conservative property of the scattering operator, we have that
\begin{equation}
\int_{4\pi}{e^{(k)}\, d\Omega}=0, \quad k\ge 1.
\end{equation}
Then taking moment of Eq. \eqref{eq:Aorder-2} we can obtain the following macroscopic equation,
\begin{equation}
\partial_{t_0} E+\nabla_0\cdot\bm{q}^{(1)}=0,
\end{equation}
where $\bm{q}^{(1)}=\int_{4\pi}{\v e^{(1)}\, d\Omega}$, which can be obtained from Eq. \eqref{eq:Aorder-1} as
\begin{eqnarray}
\bm{q}^{(1)}&=&-\tau\nabla_0 \cdot \int_{4\pi}{\v\v e^{(0)}\, d\Omega}\nonumber \\
&=&-v_g^2\tau\nabla_0 \cdot \int_{4\pi}{\s\s e^{(0)}\, d\Omega}\nonumber\\
&=&-\dfrac{v_g^2\tau}{3}\nabla_0 E,
\end{eqnarray}
which is exactly the Fourier law. Therefore, the heat transfer equation at the (macroscopic) diffusive length and time scales are
\begin{equation}
\partial_{t_0} E-\nabla_0\cdot(\kappa\nabla_0 E)=0,
\end{equation}
or
\begin{equation}
\partial_t E-\nabla\cdot(\kappa\nabla E)=0,
\end{equation}
with $\kappa=v_g^2\tau/3$ being the heat conductivity.

With the equations at different orders of $\epsilon$ given by Eq. \eqref{eq:Aorder}, we can obtain an approximation of the energy distribution function at the first order of $\epsilon$,
\begin{equation}
e''\approx e^{(0)}+\epsilon e^{(1)}=e^{eq}-\tau\epsilon\v\cdot \nabla_0 e^{eq} = e^{eq}-\tau\v\cdot \nabla e^{eq},
\end{equation}
which is used in the discussion on the AP property of the DUGKS in the diffusive limit, i.e., Eq. \eqref{eq:1st-order expansion}.

\end{document}